\documentclass[sigconf,nonacm]{cidr-2025}

\renewcommand{\paragraph}[1]{\vspace{0.2em}\noindent\textbf{#1.}}

\usepackage{enumitem}
\setlist[itemize]{nosep,leftmargin=*}
\setlist[enumerate]{nosep,leftmargin=*}

\usepackage{listings}
\newcommand{\revision}[1]{\color{black}#1 \color{black}}

\newcommand{\htedit}[1]{\color{black}#1 \color{black}}

\AtBeginDocument{%
  \providecommand\BibTeX{{%
    \normalfont B\kern-0.5em{\scshape i\kern-0.25em b}\kern-0.8em\TeX}}}

\setcopyright{acmcopyright}
\copyrightyear{2025}
\acmYear{2025}

\acmConference[CIDR '25]{15th Annual Conference on Innovative Data Systems Research}{January 19--22, 2025}{Amsterdam, The Netherlands}
\usepackage{amsmath} 
\usepackage{balance} 
\usepackage{booktabs} 
\usepackage{caption}
\usepackage{comment}
\usepackage{enumitem}
\setlist[itemize]{leftmargin=2.2em}

\usepackage{fancyvrb} 
\usepackage[symbol]{footmisc}
\usepackage{graphicx} 
\usepackage{makecell} 
\usepackage{multirow} 
\usepackage{soul}
\usepackage{subfigure} 
\usepackage{xcolor}
\usepackage{xspace} 
\usepackage{bm}
\usepackage{utfsym}
\usepackage{cancel}
\usepackage[linesnumbered,ruled,vlined]{algorithm2e}
\usepackage{pifont}
\usepackage{tikz}

\newcommand{\term}[1]{\textcolor{black}{#1}}

\theoremstyle{definition}

\theoremstyle{plain}

\theoremstyle{remark}

\newcommand\dbname{\ensuremath{\textsf{NeurDB}}\xspace}
\newcommand\dbnamecc{\ensuremath{\textsf{NeurDB(CC)}}\xspace}

\newcommand\cacheF{\ensuremath{\mathcal{{F}}}}

\newcommand\queryQ{\ensuremath{\mathcal{{Q}}}}
\makeatletter
\patchcmd{\@algocf@start}
{-1.5em}
{0pt}
{}{}
\let\oldnl\nl
\newcommand{\nolinum}{\renewcommand{\nl}{\let\nl\oldnl}}
\makeatother

\begin{document}


\title{{\dbname}: On the Design and Implementation of an
AI-powered Autonomous Database}

\author{
    Zhanhao Zhao\textsuperscript{1}, 
    Shaofeng Cai\textsuperscript{1}, 
    Haotian Gao\textsuperscript{1}, 
    Hexiang Pan\textsuperscript{1}, 
    Siqi Xiang\textsuperscript{1}, 
    Naili Xing\textsuperscript{1}, \\ 
    Gang Chen\textsuperscript{2}, 
    Beng Chin Ooi\textsuperscript{1}, 
    Yanyan Shen\textsuperscript{3}, 
    Yuncheng Wu\textsuperscript{4}, 
    Meihui Zhang\textsuperscript{5}
}
\affiliation{%
  \institution{
  \textsuperscript{1}National University of Singapore
  \textsuperscript{2}Zhejiang University
  \textsuperscript{3}Shanghai Jiao Tong University \\
  \textsuperscript{4}Renmin University of China \textsuperscript{5}Beijing Institute of Technology
  }
  \city{}
  \country{}
}
\email{{zhzhao,shaofeng,gaohaotian,xiangsq,xingnl,ooibc}@comp.nus.edu.sg, panh@u.nus.edu}
\email{cg@zju.edu.cn,shenyy@sjtu.edu.cn,wuyuncheng@ruc.edu.cn,meihui_zhang@bit.edu.cn}

\renewcommand{\shortauthors}{Zhanhao Zhao et al.}

\sloppy

\begin{abstract}
Databases are increasingly embracing AI to provide intelligent in-database analytics and autonomous system optimization, aiming to relieve end-user burdens across various industry sectors.
Nonetheless, most existing approaches fail to account for the dynamic nature of databases, which renders them ineffective for real-world applications characterized by 
\term{evolving data and workloads}.
This paper introduces \dbname, an \term{AI-powered autonomous database} that deepens the fusion of AI and databases with adaptability to \term{data and workload drift}.
\dbname establishes a new \term{in-database AI ecosystem} that seamlessly integrates AI workflows within the database.
This integration enables \term{efficient and effective in-database AI analytics} and \term{fast-adaptive learned database components}.
Empirical evaluations demonstrate that \dbname substantially outperforms existing solutions in managing AI analytics tasks, with the proposed learned components more effectively handling data and workload drift than state-of-the-art approaches.

\end{abstract}

\maketitle

\section{Introduction}


\revision{Database management systems (DBMSs) are becoming more intelligent and autonomous by embracing artificial intelligence (AI), catering to modern data- and AI-centric applications.
AI augments DBMS functionality to support in-database AI analytics~\cite{DBLP:conf/cidr/KaranasosIPSPPX20,DBLP:journals/pvldb/XingCCLOP24,DBLP:conf/mm/OoiTWWCCGLTWXZZ15,DBLP:conf/mm/WangCDGOTW15}, enabling complex and advanced analytics tasks such as disease progression predictions and user purchase recommendations.
In parallel, AI empowers DBMSs to achieve autonomous DBMS optimization, driving a broad range of learned database designs that enhance system performance with minimal human intervention, including learned knob tuners~\cite{DBLP:conf/cidr/PavloAALLMMMPQS17}, learned system drivers~\cite{DBLP:journals/pvldb/ZhuWWWPWDLZZ24,DBLP:journals/pvldb/0001ZSYHJLWL21}, 
and learned database components~\cite{DBLP:conf/cidr/KraskaABCKLMMN19} such as query optimizers~\cite{DBLP:conf/sigmod/MarcusNMTAK21,DBLP:journals/pvldb/ZhuCDCPWZ23}, indexes~\cite{DBLP:conf/sigmod/KraskaBCDP18}, and concurrency control~\cite{DBLP:conf/osdi/WangDWCW0021}, etc.}



The aspiration of integrating DBMSs with AI was first expressed forty years ago~\cite{DBLP:conf/aaai/Brodie88}, 
which has been periodically revisited with evolving technology~\cite{DBLP:journals/sigmod/0059Z0JOT16,DBLP:conf/sigmod/CongYZ24,DBLP:conf/sigmod/ReABCJKR15}.
With the advancements in both AI and DBMSs, considerable progress has been made in deepening their fusion.
Unfortunately, a huge gap remains between the potential of this integration and its current state of usability, largely due to the inherent differences in their paradigms~\cite{neurdb-scis-24}.
The dynamic nature of databases, characterized by \emph{data and workload drift}, poses a fundamental challenge~\cite{DBLP:journals/pacmmod/WuI24,DBLP:journals/pacmmod/KurmanjiTT24,DBLP:journals/pacmmod/KurmanjiT23}.
\htedit{
For example, suppose an e-commerce database stores user and product information.
While the database continuously evolves due to user and merchant activities in normal periods, it would experience a sharp increase in workloads during flash sales, where huge volumes of updates from sales transactions are processed.
}
In contrast, AI models typically derive intelligence from static datasets and thus can become outdated quickly in the face of database dynamism.
For instance, models for user purchase recommendations typically acquire knowledge by training on specific user behaviors and product information.
However, when the data drifts due to transactional updates, models relying on past knowledge may 
\htedit{produce}
inaccurate recommendations since they are not updated to reflect new conditions.
Similarly, learned query optimizers trained based on historical system environments including data distributions and workload patterns 
\htedit{can}
struggle to deliver effective query plans as continuous data and workload drift.





\revision{
Adapting to data and workload drift has gained traction for both in-database AI analytics and autonomous DBMS optimization.
Existing databases supporting in-database AI analytics~\cite{azure,postgresml,mindsdb} often recommend users to completely retrain models with new data and workloads when the models
become outdated.
This retraining process is typically performed manually outside the database, 
and therefore, it may complicate 
the AI analytics workflow and becomes inefficient when dealing with continuous data and workload drift.
%
\htedit{To tackle this problem,} in the context of autonomous DBMS optimization, recent works~\cite{DBLP:journals/pacmmod/WuI24,DBLP:conf/sigmod/LiLK22,DBLP:journals/pvldb/LiWZDZ023,DBLP:journals/pacmmod/KurmanjiT23} have started addressing the adaptability issues by automatically detecting drift and triggering model retraining.
However, these approaches are often tailored to certain components.
For example, 
approaches~\cite{DBLP:journals/pacmmod/WuI24,DBLP:journals/pacmmod/KurmanjiT23} 
\htedit{that allow} 
learned query optimizers to withstand workload pattern changes can be ineffective for learned indexes requiring rapid adaptation to data distribution drift.
Consequently, a paradigm that robustly handles data and workload drift for various learned database components remains absent.
}





We envision a deeper integration of AI and DBMSs for enhanced adaptability to both data and workload drift.
Unlike existing approaches that merely overlay AI onto DBMSs or selectively enhance certain system components, we 
aim for 
a comprehensive fusion of AI with DBMSs that enables continuous adaptation to drift.
It promises a cohesive AI-powered DBMS that provides advanced in-database AI analytics and deeply integrates AI into all its key components.
However, achieving this level of integration presents significant challenges.
First, 
the integration 
necessitates
a fundamental redesign of the entire AI workflow, including training, inference, and fine-tuning, within the database architecture.
Second, models for AI analytics and learned database components must swiftly adapt to data and workload drift without losing effectiveness and efficiency.
Therefore, developing such an AI-powered database requires both a \term{system foundation that seamlessly supports model adaptation} and \term{AI models that are inherently adaptive in structure}.



In this paper, we present \dbname, an AI-powered autonomous DBMS that provides efficient and effective in-database AI analytics to seamlessly support modern AI-powered applications, and fully embraces AI techniques in each major system component to offer autonomous system optimization.
At the core of \dbname is an in-database AI ecosystem that 
\htedit{deeply integrate the AI workflow into the database.}
In this ecosystem, we develop multiple in-database AI operators, such as model training, inference, and fine-tuning, along with an in-database AI engine to handle the execution of these operators.
\dbname can then directly support in-database AI analytics by calling these AI operators.
To further simplify end-users in submitting their AI analytics tasks to \dbname, we provide a user-friendly interface by extending SQL with \texttt{PREDICT} syntax.
Moreover, we propose two techniques to optimize the AI ecosystem performance.
First, we devise a data streaming protocol that significantly reduces data transfer overhead, yielding better performance for AI operations.
Second, we develop an incremental update technique to minimize the fine-tuning cost, facilitating the fast adaptation of AI models.
Based on 
the in-database AI ecosystem,
we enable two efficient learned database components, namely a learned concurrency control algorithm and a learned query optimizer, that can adapt quickly to data and workload drift.




\revision{The remainder of the paper is structured as follows.
The next section provides the necessary preliminaries and further clarifies the motivation.
Section~\ref{sec:overview} describes the system overview of \dbname, and Section~\ref {sec:design} details the key system design, including the in-database AI ecosystem and fast-adaptive learned database components.
Section~\ref{sec:evaluation} presents the experimental results.
Section~\ref{sec:related} discusses the related works before Section~\ref{sec:conclusion} concludes.
}




\section{Preliminaries and Motivation}\label{sec:preliminaries}

In this section, we introduce the preliminaries of AI and DBMS integration, outline the design goals of \dbname, and present the enhanced SQL syntax designed for in-database AI analytics.

\subsection{AI and DBMS Integration} \label{sec:compatitors}

\revision{
The integration of AI and DBMSs can offer each other mutual benefits~\cite{DBLP:journals/sigmod/0059Z0JOT16}, and building on this intuition, various research works on AI and DBMS integration have been proposed, which can be broadly categorized into two areas: in-database analytics and autonomous system optimization.
Modern databases, such as Oracle~\cite{oracle}, Microsoft Azure SQL~\cite{azure}, Amazon Redshift~\cite{amazon}, and Google BigQuery~\cite{bigquery}, support in-database AI analytics, and allow AI tasks to be performed directly through user-friendly SQL interfaces. 
For better performance, these databases enable AI-driven knob tuning to autonomously refine database configurations.

Adapting to data and workload drift has been recognized as a critical challenge in advancing the integration of AI and DBMSs~\cite{DBLP:journals/pacmmod/WuI24,DBLP:journals/pacmmod/KurmanjiTT24,DBLP:journals/pacmmod/KurmanjiT23}.
Many existing approaches for whether in-database analytics~\cite{DBLP:conf/cidr/KaranasosIPSPPX20,DBLP:journals/pvldb/SalazarDiazGR24} or autonomous system optimization~\cite{DBLP:journals/pvldb/MarcusNMZAKPT19,DBLP:conf/sigmod/KraskaBCDP18} depend on static models or complete retraining, 
\htedit{limiting their ability to handle continuous data and workload drift.}
More importantly, these approaches are typically layered on top of DBMSs rather than being holistically embedded within DBMSs.
This separation hinders real-time adaptability, because internal and fine-grained performance metrics, which are essential for detecting and addressing drift, can be difficult to capture externally~\cite{DBLP:conf/osdi/WangDWCW0021,DBLP:conf/sigmod/LiLK22}.


To mitigate these limitations, we design and develop \dbname, an AI-powered database with AI holistically embedded from the ground up.
By deeply integrating AI into all its key components and functionalities, \dbname aims to automatically manage adaptation workflows for both in-database AI analytics and autonomous system optimization.
}


\vspace{-0.15em}
\subsection{Design Goals}\label{subsec:design_goals}

DBMSs are dynamic as data and workloads evolve over time, and therefore, 
the system must 
be designed
for {\em adaptability} while 
also
guaranteeing {\em reliability} and {\em scalability}.
Building upon this understanding,
we equip \dbname with these three key properties.

\noindent\textbf{Adaptability} is the 
capability
of a DBMS to evolve autonomously in response to drifting data and workloads.
With optimal adaptability, DBMS can respond to drift in real time.

\noindent\textbf{Reliability} depicts the ability of a DBMS to consistently meet performance and accuracy standards, even during 
phases of
adaptation and evolution.
With optimal reliability, the system can 
operate at peak performance and maintain high accuracy consistently.

\noindent\textbf{Scalability} refers to the system's 
capacity
to maintain or 
enhance
performance as the workload 
increases by introducing more resources, such as threads or nodes.

\noindent Informally, 
\htedit{our design goal is}
to ensure \dbname can uphold reliability as quickly as possible when the system adapts or evolves due to data and workload drift, while ensuring good scalability.
In addition, we design \dbname as a general-purpose AI-powered DBMS which is expected to serve as a foundational infrastructure for advanced data- and AI-centric applications to enrich various domains with AI capabilities~\cite{neurdb-scis-24}.

\vspace{-1mm}
\subsection{SQL Syntax for AI Analytics} \label{subsec:sql}

\dbname incorporates enhanced SQL to support AI analytics.
As illustrated in Listing \ref{lst:syntax-reg} and Listing \ref{lst:syntax-cls}, it extends from the standard SQL by introducing a \texttt{PREDICT} keyword to handle two typical AI tasks: regression with the \texttt{VALUE OF} clause and classification with the \texttt{CLASS OF} clause. 
Inspired by the original principle of SQL that allows an application developer to write a query with \texttt{SELECT}, and then let the DBMS find the most efficient way to store and retrieve data, we ensure that a developer can submit an AI analytics task simply with \texttt{PREDICT}. 
All the following operations, such as retrieving training data and invoking AI models, are handled automatically by \dbname.
Consequently, unlike existing solutions~\cite{postgresml,bigquery} that require application developers to intervene in the execution of AI analytics, such as specifying model parameters, our 
\htedit{approach}
can make the execution transparent to application developers, relieving them from the complexities of performing 
\htedit{such}
analytics.
Further, we plan to expose additional SQL for AI model management and other AI analytics functionalities, such as generative AI. 

We now present two real-world analytics scenarios that can be directly supported by \dbname using \texttt{PREDICT} \htedit{queries.}



\paragraph{Regression} Listing \ref{lst:syntax-reg} shows the SQL to predict and fill in the missing scores of products based on user reviews.
By specifying the features for training with \texttt{TRAIN ON} and using data stored in the `review' table, we predict the value of the `score' variable.
Notably, 
\htedit{for \texttt{TRAIN ON}, the asterisk}
automatically excludes features with unique constraints to avoid including meaningless data.

\begin{lstlisting}[
    caption={Syntax for a regression task},
    label={lst:syntax-reg},
]
PREDICT VALUE OF score
FROM review
WHERE brand_name = 'Special Goods'
TRAIN ON * 
    WITH brand_name <> 'Special Goods'
\end{lstlisting}

\paragraph{Classification} The disease progression prediction in healthcare can be handed by the SQL shown in Listing \ref{lst:syntax-cls}.
Similarly, this query specifies the features for training with \texttt{TRAIN ON} and the data table using \texttt{FROM} to predict the class of the `outcome' variable, which indicates whether a patient has diabetes or not.
In addition, we support directly inputting missing data using \texttt{VALUES}.

\begin{lstlisting}[
    caption={Syntax for a classification task},
    label={lst:syntax-cls}
]
PREDICT CLASS OF outcome 
FROM diabetes
TRAIN ON pregnancies, glucose, blood_pressure, ...
VALUES (6, 148, 72, ...), (1, 85, 66, ...), ...
\end{lstlisting}

\section{System Overview}
\label{sec:overview}

\begin{figure}
\centering \includegraphics[width=\linewidth]{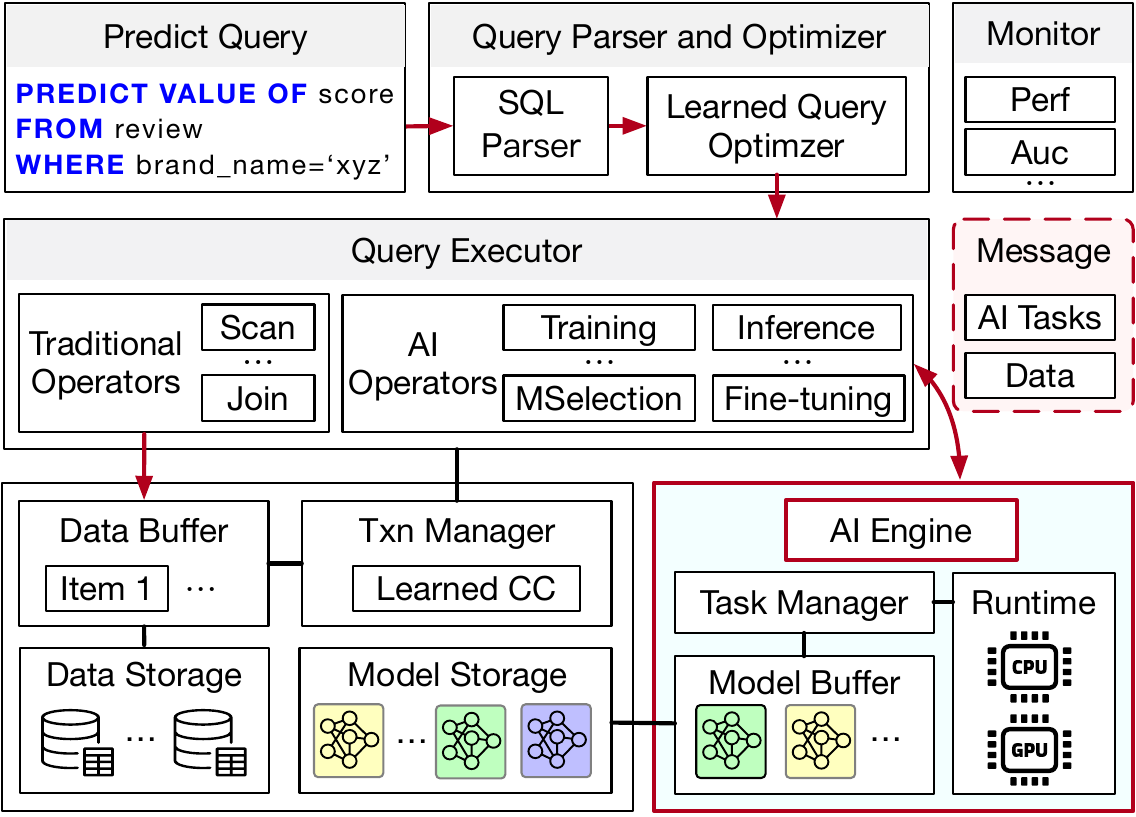} 
\caption{The System Architecture of {\dbname}} 
\label{fig:architecture} 
\vspace{-3mm}
\end{figure}

We now describe the system architecture of \dbname as shown in Figure~\ref{fig:architecture},
where we achieve the deep fusion of AI and databases by establishing an in-database AI ecosystem.
Based on this AI ecosystem, we enable efficient and effective in-database AI
analytics and develop fast-adaptive learned database components.


\paragraph{In-database AI Ecosystem}
\htedit{We establish the in-database AI ecosystem by holistically redesigning existing database components, such as the query executor, and introducing additional modules, including an AI engine and AI model storage.}
We first extend the query executor to support AI workflow.
Beyond traditional operators for fetching and processing data, e.g., scan and join, the query executor in \dbname includes in-database AI operators for model training, inference, and fine-tuning.
With these AI operators, we effectively integrate the AI workflow into the database query processing.
We are developing additional AI operators, such as model selection~\cite{DBLP:journals/pvldb/XingCCLOP24} and model slicing~\cite{DBLP:journals/pvldb/zengslicing25}, to provide more comprehensive AI services.
For example, a query may call the model selection operator (denoted as MSelection) to automatically select the best-suited model for a given prediction task, thereby enhancing accuracy and efficiency.
We then introduce a new module, called the {\em AI engine}, into \dbname.
It handles the AI-related processing requests from AI operators and learned components and creates AI tasks on their behalf, 
where the task manager dispatches them to the CPU/GPU runtime for execution. 
Specifically, we propose a data streaming protocol 
to reduce data transfer overhead, and an incremental model update technique to facilitate fast model adaptation, thus enhancing the performance of the AI engine.
We shall elaborate on these optimizations in Section~\ref{subsec:ai_design}.
Also, we design dedicated {\em AI model storage} to store AI models and serve them based on requests from the AI engine.
We further implement a monitor to detect unexpected performance or accuracy issues, based on which we trigger automatic and appropriate model adaptation.


\paragraph{In-database AI Analytics}
In Figure~\ref{fig:architecture}, we illustrate a running example of a \texttt{PREDICT} query.
We enhance the SQL parser and the query optimizer to produce a customized query plan for \texttt{PREDICT} queries.
After parsing and optimizing the query, the query executor then executes the query according to the query plan.
In particular, it performs the scan operator to retrieve data and then invokes the inference operator to deliver the inference task to the AI engine.
Next, the AI engine uploads the model from the model storage if it is not in the model buffer, and conducts model inference to produce the results.
If the model is detected to be inaccurate, \dbname invokes the fine-tuning operator to update the outdated model with the help of the AI engine. 


\revision{
\paragraph{Autonomous DBMS Optimization}
We also enable autonomous system optimization based on the proposed in-database AI ecosystem.
For instance, the learned query optimizer interacts with the AI engine to train a model responsible for generating efficient query plans. 
The monitor tracks the performance of the generated query plans, and if a plan is identified as inefficient, the monitor notifies the AI engine to fine-tune the model, enabling continuous optimization and adaptability.
Similarly, the monitor can trigger autonomous knob tuning when suboptimal knob settings are detected, ensuring that the system remains well-configured to handle data and workload drift effectively.
We will further introduce the detailed design of learned database components in Section~\ref{sec:learned_components}.
}

\section{The Design of {\dbname}} \label{sec:design}

In this section, we present 
two key subsystems of {\dbname}, namely an in-database AI ecosystem and learned database components.

\subsection{In-database AI Ecosystem}\label{subsec:ai_design}

\begin{figure}[t]
    \centering
    \includegraphics[width=\linewidth]{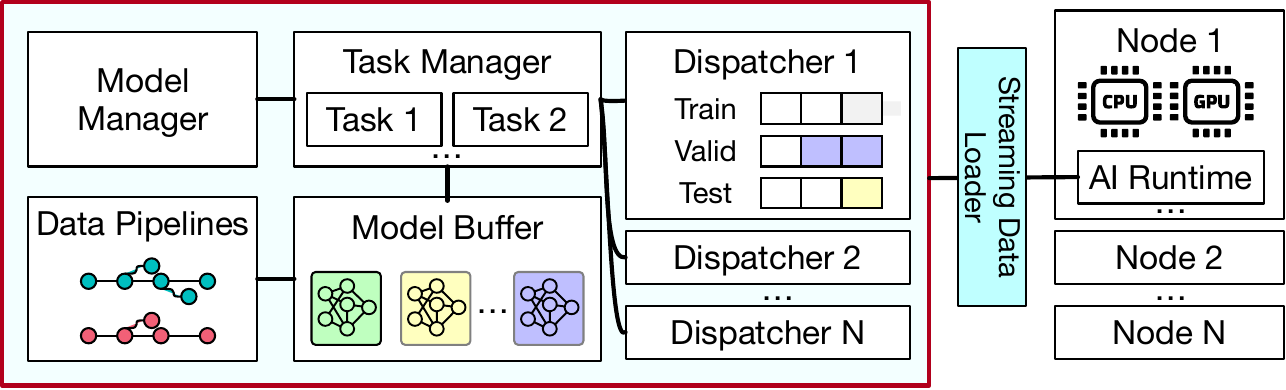}
    \caption{AI Engine of \dbname}
    \label{fig:ai-engine}
    \vspace{-3mm}
\end{figure}

\paragraph{AI Engine}
%
The AI engine of \dbname, pivotal to all AI-related activities for tasks from both user and internal learned components, operates on a distributed and event-driven architecture to optimize efficiency and throughput.
Figure~\ref{fig:ai-engine} shows the main components and the communicative flow with connected external nodes serving distributed AI tasks. 
In the AI engine, the {\em task manager} is the main component that coordinates and schedules the tasks and resources.
It handles and parses the incoming AI tasks, and creates a {\em dispatcher} for each task.
A dispatcher connects to multiple AI runtimes at external nodes.
It also loads and caches the necessary data required by the corresponding AI tasks, 
performs data pipelines on it for preprocessing, feature engineering, etc, and pushes the prepared data and model weights to the remote AI runtime to trigger AI activities.
Notably, the data is transferred in a streaming and pipelining manner to minimize the delay in the data preparation steps.

\paragraph{Data Streaming Protocol}
During an AI task, the AI runtime receives data continuously from the database.
\dbname's AI engine optimizes this process with a dedicated data streaming protocol to reduce the time and memory overheads.
Specifically, the AI runtime establishes a TCP socket connection with the dispatcher.
When a task is assigned to the dispatcher, it first schedules the AI runtimes and performs handshakes with them to negotiate (1) model parameters, such as model structure, model arguments, the training batch size, etc, and (2) streaming parameters, e.g., the initial size for send and receive buffers and the number of batches per transmission.
Then, it starts the data and model transfer through the connection.
Notably, to support adaptable control and scheduling over resources, these parameters can be dynamically updated for an ongoing AI task through a data-driven dispatcher.
This makes the AI engine a self-driving component, 
thereby controlling and optimizing its operations autonomously.

\paragraph{Model Manager}
%
Given the dynamic nature of DBMSs, 
a typical AI lifecycle extends beyond a single model.
%
As new data is introduced, incremental updates and retraining are required to address data and workload drift that degrade predictive performance.
%
This process results in multiple evolving model versions, creating significant management difficulties and storage overheads.
To address this challenge, \dbname leverages the capabilities of databases to efficiently manage AI models created by either users or internal components and handle drift by design.
Specifically, it introduces a dedicated {\em model manager} in its AI engine, enabling fine-grained model management with efficient updates.
%
The model manager provides high-level interfaces for handling AI operations, such as training, inference, and fine-tuning, executed via {\em model views}.
%
Similar to data views in DBMSs, model views serve as logical abstractions of AI models tailored for specific tasks, with physical representations maintained in model storage.
Formally, a model $M$ 
%
comprises a series of layers $L^{(j)} (j \in \mathbb{Z}^{+})$.
To generate outputs for data $X$, the layers are executed sequentially, i.e., 
$M(X) = L^{(n)}(L^{(n-1)}(...(L^{(1)}(X)))$.
%
This layered model storage approach aligns with the structure of deep neural networks (DNNs), ensuring fast and efficient AI model accessibility.
%
Notably, this model formulation can also be generalized to non-linear DNN architectures that are represented as directed acyclic graphs, which can be achieved by executing layers based on the topological order.


%

\begin{figure}[t]
    \centering
    \includegraphics[width=\linewidth]{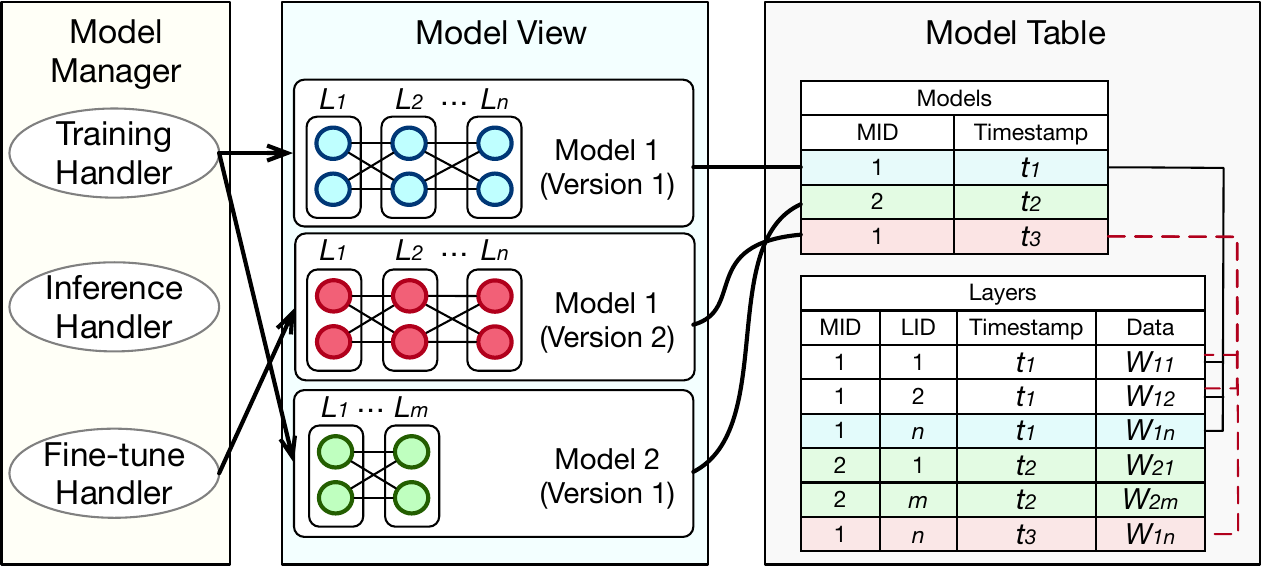}
    \caption{Incremental Update for Model Manager}
    \label{fig:model-update}
    \vspace{-3mm}
\end{figure}

\paragraph{Model Incremental Update}
%
Leveraging its layered model storage, \dbname enables versioning for a specific model and supports incremental model updates through fine-tuning.
%
In particular, to adapt the current model to drifting data distributions, the AI engine selectively fine-tunes the final layers using the updated data stored in the designated database relation and freezes the preceding layers.
%
Subsequently, only the updated layers remain persistent in the model storage, from where they can be extracted and merged with the previously frozen layers to create a new model version.

Figure~\ref{fig:model-update} illustrates how the model manager supports incremental updates for AI models.
%
%
%
%
Whenever
an
inference or fine-tuning task is 
initiated,
the system retrieves all layers up to the most recent version to construct the model,
i.e., for a model $M_{i, t}$, uniquely defined by its Model ID (MID) $i$ and the timestamp of creation $t$, consisting of $k$ layers,
\begin{gather*}
M_{i, t}(X) = 
    L_{i, t_k}^{(k)}(
        L_{i, t_{k-1}}^{(k-1)}(
            ...(
                L_{i, t_1}^{(1)}(X)
            )
        )
    )
\\
\;s.t.\; 
\forall p, q \in \{1, ..., k\}, 
L_{i, t_p}^{(p)} 
\rightarrow 
     t_p \geq t_q \wedge t_p \leq t .
\end{gather*}
Let us consider the example shown in Figure~\ref{fig:model-update}. 
Supposing the model $M_{1}$ requires fine-tuning the last layer $L_n$, $M_{1, 2}$, namely the second version of $M_{1}$, can then be assembled by layers
$\{L_{1, 1}^{(1)}, ..., L_{1, 1}^{(n-1)}, L_{1, 2}^{(n)}\}$.
This allows $M_{1, 1}$ and $M_{1, 2}$ to share the majority of model weights, ensuring adaptability 
to
data drift while 
maintaining efficient storage.

\subsection{Fast-adaptive Learned Components} \label{sec:learned_components}

In this section, we introduce two key learned database components, 
learned concurrency control and learned query optimizer, which can achieve fast adaptation to data and workload drift.
To achieve this, we fundamentally rely on the underlying AI ecosystem.
We non-intrusively monitor the system conditions such as transaction throughput and data distributions, which can be used to detect data and workload drift in real time.
Based on these metrics, \dbname automatically triggers model fine-tuning to adapt to continuously evolving data and workloads, and continually generates valid input for model pre-training, allowing the model for learned database components to gain global knowledge of most drift.

\begin{figure}[t]
\centering
\includegraphics[width=\linewidth]{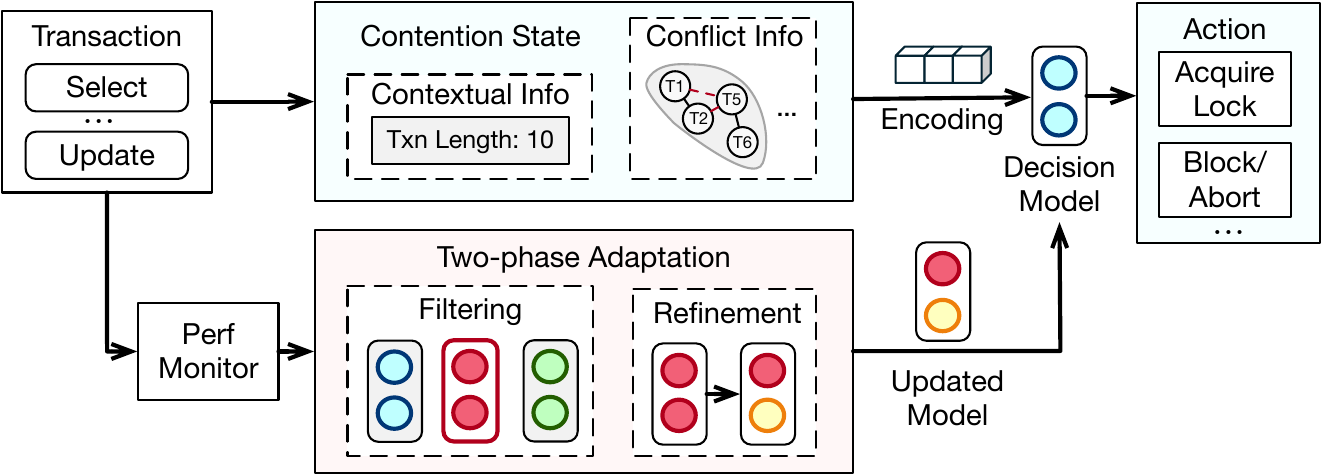}
\caption{Fast-adaptive Learned Concurrency Control}
\label{fig:learned_cc}
\vspace{-3mm}
\end{figure}

\paragraph{Learned Concurrency Control} \label{subsec:transaction_model}
We design an efficient learned concurrency control algorithm that can continually adapt to changing workloads, as shown in Figure~\ref{fig:learned_cc}.
Given a transaction $T$ with multiple operations, our algorithm incorporates a decision-making model {\cacheF} to assign each operation $op$ the optimal concurrency control action $\delta$ based on the current system condition represented as $x$, i.e., $\delta = \cacheF(x)$.
Unlike state-of-the-art approach~\cite{DBLP:conf/osdi/WangDWCW0021} that simply adjusts actions based on predefined transaction or operation patterns (e.g., transaction type), our approach learns the optimal action based on the contention state, which includes both conflict information (such as dependency) of transactions and contextual information (such as the transaction length).
For instance, 
%
%
%
%
when a write is performed on high-contention data records, we may immediately abort the transaction to avoid unnecessary costs, as the transaction is likely to be aborted eventually.
In contrast, we may execute a read on low-contention records with optimistic concurrency control without locking to avoid extra conflict detection overhead, and a long transaction that is supposed to run for a long time should have high priority. 
By using a modeling paradigm based on the contention state, our model is more likely to generalize to drifting workloads with varying levels of contention.
However, since transactions can be completed in milliseconds, the model must be efficient so as not to become a bottleneck.
To achieve this, we first develop a fast encoding technique to significantly reduce the dimension of contention state representation, and then compress the model with a flattened layer to improve inference efficiency.


As the model is compressed to trade generalizability for performance, the representation of the contention state may gradually become inaccurate over time due to workload drift.
Consequently, when we detect workload drift through their impact on transaction performance, we introduce a fine-tuning process for {\cacheF} to adapt it into $\mathcal{F}_{next}$. 
With the leaner architecture of the model, the adaptation can be accelerated due to the narrower search space compared to a large model.
Specifically, we propose a two-phase adaptation algorithm based on the online Reinforcement Learning (RL) framework.
In the first filtering phase, we generate several improved models using Bayesian optimization and evaluate them over a specific timeframe to identify the best-performing model. 
Then, in the refinement phase, we employ reward-based feedback to further optimize the selected model.

\begin{figure}[t]
\centering
\includegraphics[width=\linewidth]{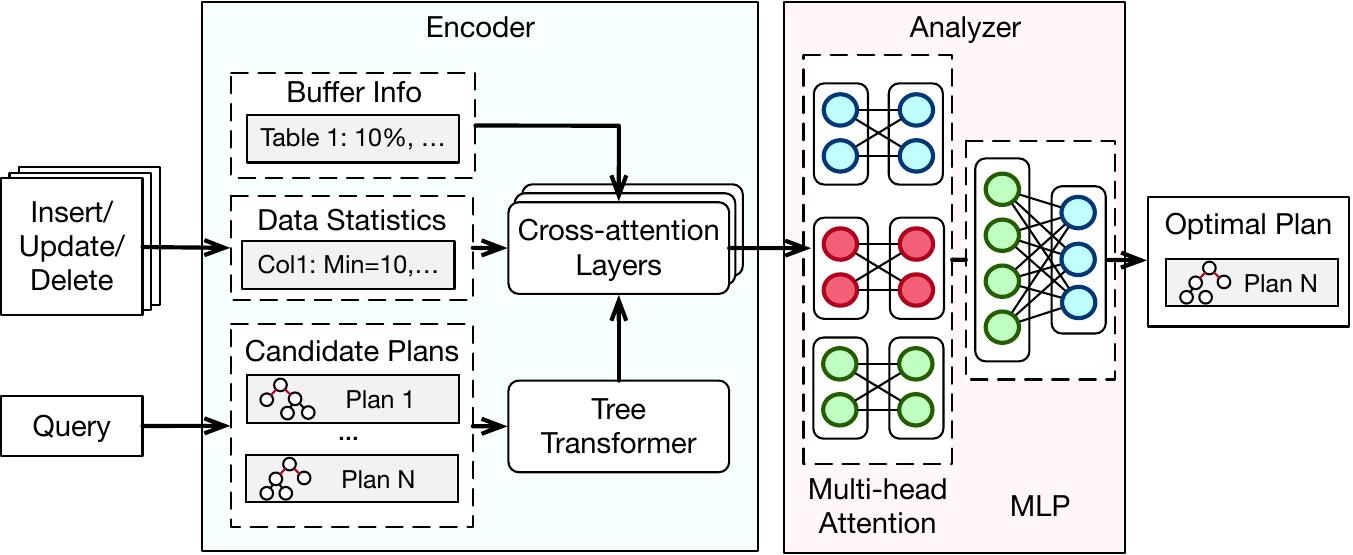}
\caption{Fast-adaptive Learned Query Optimizer}
\label{fig:learned_query_optimizer}
\vspace{-3mm}
\end{figure}

\paragraph{Learned Query Optimizer} \label{subsec:optimizer_model}
We propose a learned query optimizer that can efficiently adapt to data and workload drift, as shown in Figure~\ref{fig:learned_query_optimizer}.
In contrast to
existing works~\cite{DBLP:conf/sigmod/MarcusNMTAK21,DBLP:journals/pvldb/ZhuCDCPWZ23}
that
attempt to produce the best plan for a given query $\mathcal{Q}$ under fixed system conditions (i.e., data distribution and workload), our approach effectively identifies the plan best suited for the current system conditions.
To achieve this, we design a dual-module model consisting of an encoder and an analyzer.
Specifically, in the encoder, we input the vector generated by a tree transformer, which includes multiple candidate query plans for \queryQ, along with system condition representations (including buffer information depicting buffer usage and data statistics representing each attribute's distribution), into cross-attention layers to generate a unified embedding. 
The analyzer then uses this embedding as input for a multi-head attention layer followed by a multilayer perceptron (MLP) to deliver the optimal plan.
With more knowledge of the mapping between plans and system conditions, the model can better generalize to data and workload drift.
To maximize this knowledge, we generate various synthetic data distributions and workloads using Bayesian optimization, and pre-train
the model to handle most drift effectively.
Consequently, the proposed learned query optimizer provides consistent query performance under evolving data and workloads.
 
%

\revision{
\paragraph{Discussion}
The main design principle we follow to develop computational models and algorithms in \dbname is the filter-and-refine principle (FRP)~\cite{neurdb-scis-24,DBLP:conf/ideas/OoiPWWY02,frp}.
FRP employs a two-stage strategy: a filtering stage quickly eliminates less promising or irrelevant objects (e.g., data, strategies, etc.) from a large set, followed by a more resource-intensive refinement stage on the filtered subset.
This strategy enables the efficient execution of computationally demanding tasks while yielding near-optimal results.
We therefore leverage FRP in both in-database AI analytics and autonomous DBMS optimization for enhanced adaptability.
For example, we integrate FRP in the RL framework of learned concurrency control and incorporate it in the multi-head attention layers of the learned query optimizer.
However, existing model structures, typically characterized by fixed topologies and layer configurations, constrain the full potential of FRP.
We are exploring a flexible model structure that can be dynamically adjusted with the guidance of FRP, enabling seamless adaptation to data and workload drift.
}

\section{Evaluation} \label{sec:evaluation}

In this section, we present our preliminary evaluation results. 
We first introduce the experimental setup, and then evaluate the in-database AI ecosystem under real-world AI analytics scenarios, and the learned database components with data and workload drifts.

\subsection{Experimental Setup}
We conduct experiments on a server equipped with an Intel(R) Xeon(R) W-2133 CPU@3.60GHz (12 cores, 24 threads), 64 GB memory, and 3 NVIDIA GeForce RTX 2080 Ti GPUs.
All experiments are executed within Docker containers based on the official Ubuntu 22.04 image with CUDA 11.8.0, leveraging the host's GPU resources.

\subsubsection{Benchmarks}
We construct two real-world applications that 
require
AI analytics.
The AI analytics queries used in our experiments are listed in Table~\ref{table:ai_queries}.
\noindent
\begin{itemize}[nosep,leftmargin=*]
    \item \textbf{E-commerce (E) Workload} performs click-through rate prediction, a critical task in e-commerce for product recommendations, using the Avazu dataset (Avazu)~\cite{avazu}, which consists of \textasciitilde40.4M records and 22 attributes.
    We use k-means clustering to create five data clusters namely, $C_1$ to $C_5$, and by switching from one to another, we simulate the data distribution drift.

    \item \textbf{Healthcare (H) Workload} conducts disease progression prediction using the UCI Diabetes dataset (Diabetes)~\cite{diabetes}.
    After scaling, the dataset comprises \textasciitilde5.2M data records and 43 attributes. 
\end{itemize}



\vspace{0.5mm}
\noindent We establish a \textbf{micro-benchmark} to evaluate the learned database components.
It consists of a transactional benchmark based on YCSB~\cite{DBLP:conf/cloud/CooperSTRS10}, which generates synthetic workloads for large-scale Internet applications.
Each transaction performs 5 selects and 5 updates on a table with 1 million records.
In addition, we construct an OLAP benchmark based on the STATS dataset~\cite{stats}, which consists of 8 tables from the Stats Stack Exchange network.
We execute inserts/updates/deletes with randomly generated data values to simulate data distribution drift following a recent work~\cite{DBLP:journals/pvldb/LiWZDZ023}.

\begin{table}[t]
\setlength{\abovecaptionskip}{0.1cm}
\caption{Queries for AI Analytics Evaluations}
\resizebox{0.95\columnwidth}{!}{
\renewcommand\arraystretch{1.2}{
\begin{tabular}{ll}
\toprule
\textbf{Workload}     & \textbf{Statement}
\\ \midrule
E-Commerce (E)      & \texttt{PREDICT VALUE OF} click\_rate \texttt{FROM} avazu \texttt{TRAIN ON} *                        \\ 
Healthcare (H)     & \texttt{PREDICT CLASS OF} outcome \texttt{FROM} diabetes \texttt{TRAIN ON} *                        \\ 
\bottomrule
\end{tabular}
}
}
\vspace{-0.5em}
\label{table:ai_queries}
\end{table}






\subsubsection{Implementation and Default Configuration}
We have released the first version of \dbname 
\cite{neurdbcode},
in which
we integrate our proposed in-database AI ecosystem and learned components into the codebase of PostgreSQL v16.3.  
We will gradually introduce new modules such as AI-powered resource scheduling, etc., and replace existing components where necessary.
We implement a baseline system called PostgreSQL+P, which loads data from PostgreSQL in batches, and utilizes an AI runtime built with PyTorch to support AI analytics.
By default, we employ ARM-Net~\cite{cai2021arm}, an adaptive relation modeling network tailored for structured data, as the basic analytics model for both PostgreSQL+P and \dbname.
In our experiments, we inherit the default settings of PostgreSQL unless otherwise specified.
We set the default window size of the streaming data loader to 80 data batches.
Each batch contains 4096 data records (samples).


\begin{figure}[t]
\centering
\subfigure[Overall Performance]{
\begin{minipage}[t]{0.48\linewidth}
\label{Fig.indb_analytics.throughput}
\includegraphics[width=1\textwidth]{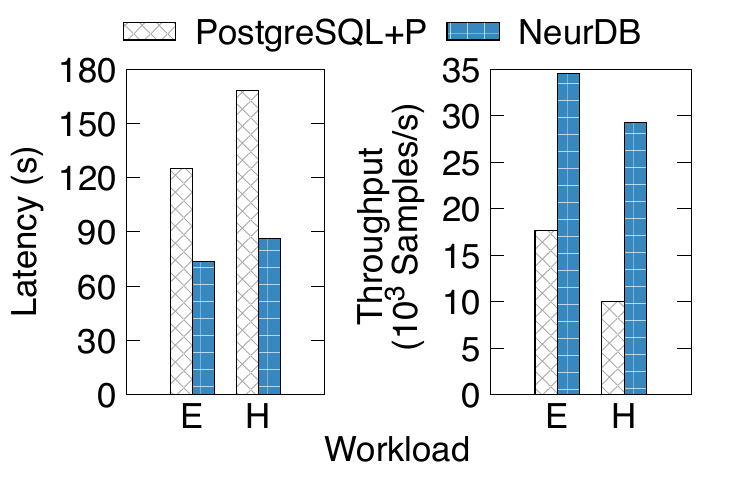}
\end{minipage}%
}
\subfigure[Effects of Data Volume]{
\begin{minipage}[t]{0.48\linewidth}
\label{Fig.indb_analytics.latency}
\includegraphics[width=1\textwidth]{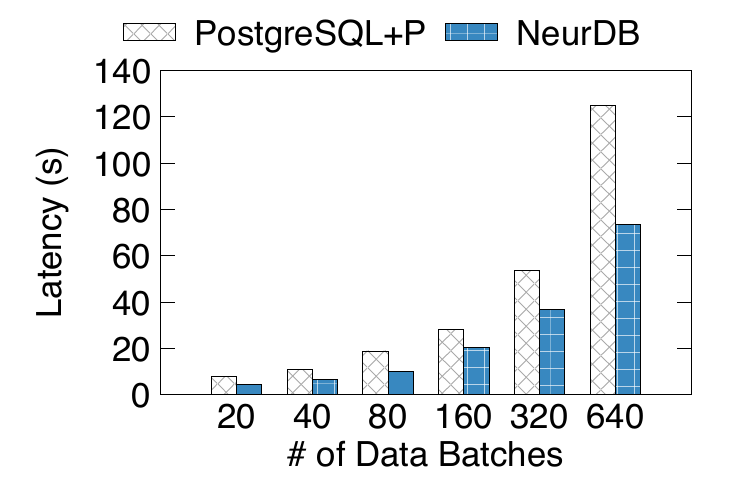}
\end{minipage}%
}
\subfigure[Effects of Data Distribution Drift]{
\begin{minipage}[t]{0.95\linewidth}
\label{Fig.indb_analytics.finetune}
\includegraphics[width=1\textwidth]{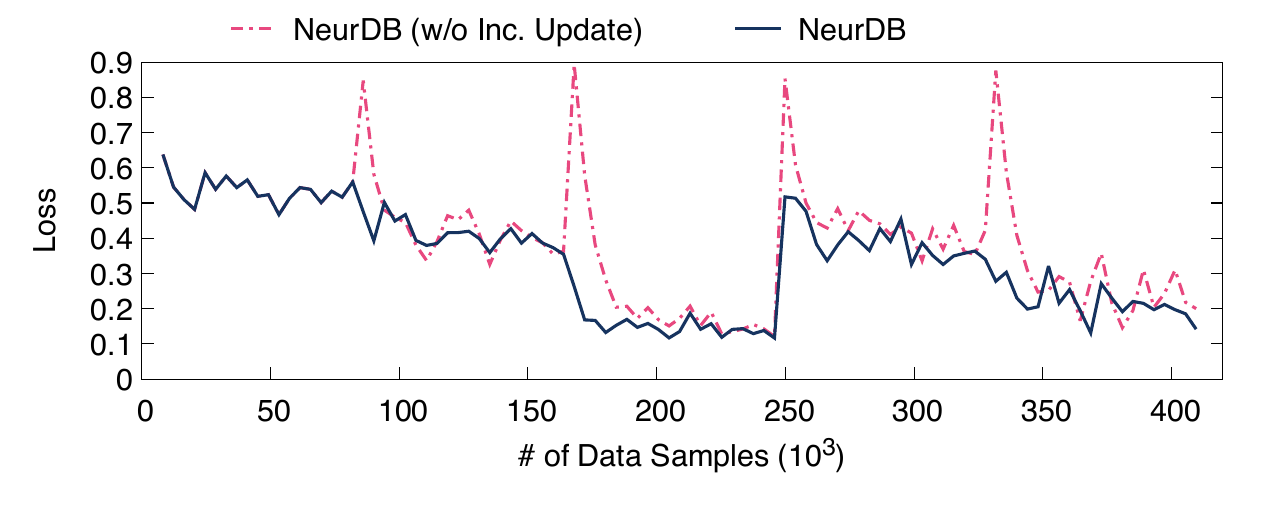}
\end{minipage}%
}
\captionsetup{justification=raggedright}
\vspace{-2mm}
\caption{End-to-end Performance of AI Analytics}
\label{fig.indb_analytics_perf}
\end{figure}

\subsection{In-database AI Analytics}

We evaluate the in-database AI analytics in terms of end-to-end latency, training throughput, and loss variation with data drift.

\paragraph{Efficiency and Scalability}
%
We study the efficiency of \dbname on AI analytics workloads by comparing it with PostgreSQL+P.
As observed in Figure~\ref{Fig.indb_analytics.throughput}, \dbname achieves up to 41.3\% and 48.6\% lower end-to-end latency, and 1.96$\times$ and 2.92$\times$ higher training throughput than PostgreSQL+P for Workload E and Workload H, respectively. 
The significant improvement in latency and throughput achieved by \dbname is due to its in-database AI ecosystem, which efficiently supports AI analytics by utilizing a data streaming protocol.
Further, to evaluate the impact of data volume on the end-to-end latency, we run Workload E with varying numbers of data batches.
As shown in Figure~\ref{Fig.indb_analytics.latency}, \dbname consistently outperforms PostgreSQL+P, indicating that \dbname can scale well with increased data volume.

\paragraph{Adaptability}
We investigate \dbname's ability to adapt to the drifting data and workloads with the model incremental update technique.
To simulate such drift, for $i \in [1, 4]$, we let \dbname perform the training task using cluster $C_i$ of Workload E, and 
switch to $C_{i+1}$ when 81,920 samples of $C_i$ are consumed by model training.
%
Figure~\ref{Fig.indb_analytics.finetune} plots the training losses with and without model incremental updates.
From the result, 
we can observe that starting from the first data drift, the AI engine equipped with incremental updates receives lower loss values during the sudden drift in
data distributions.
This enables the model to converge faster,
and as a result,
\dbname is equipped to
serve the new tasks effectively.


\subsection{Learned Database Components}
We now investigate the performance of our proposed learned database components.

\begin{figure}[t]
\centering
\includegraphics[width=0.95\columnwidth]{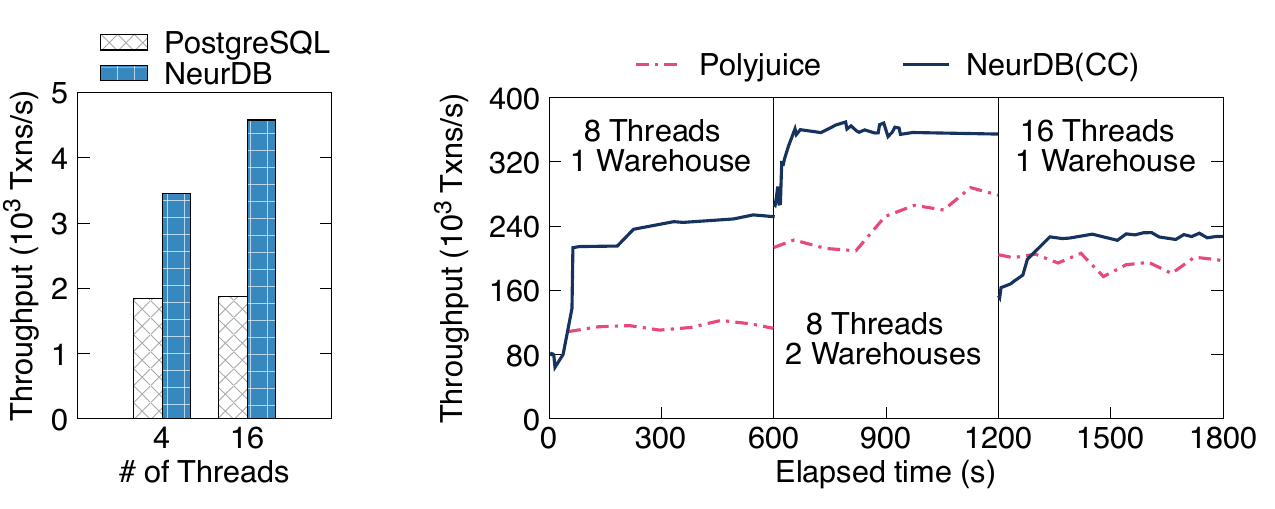}
\begin{minipage}{\columnwidth}
\vspace{-3mm}
\footnotesize
\hspace{3mm} (a) Overall Throughput 
\hspace{3mm} (b) Throughput under Data and Workload Drift
\end{minipage}
\captionsetup{justification=raggedright}
\caption{Performance of Learned Concurrency Control}
\label{fig.learned_cc_perf}
\vspace{-1mm}
\end{figure}

\paragraph{Learned Concurrency Control} 
To evaluate the proposed concurrency control algorithm, we compare \dbname with PostgreSQL using the micro-benchmark with varying thread counts.
The results, shown in Figure~\ref{fig.learned_cc_perf}(a), 
demonstrate that \dbname achieves up to 1.44$\times$ higher transaction throughput than PostgreSQL.
We attribute this performance gain to the proposed algorithm's ability to schedule transactions more effectively than serializable snapshot isolation~\cite{DBLP:journals/pvldb/PortsG12}, a static concurrency control algorithm employed in PostgreSQL.
We further evaluate the adaptability of our proposed algorithm 
against
Polyjuice~\cite{DBLP:conf/osdi/WangDWCW0021}, state-of-the-art learned algorithm.
Due to the limitation of constraining the transaction execution workflow, implementing Polyjuice on the codebase of \dbname 
would be cumbersome.
We therefore opt to implement our algorithm, named \dbnamecc, into the Polyjuice codebase to facilitate a fair comparison.
We set up a drift workload based on TPCC~\cite{tpcc} by varying the number of warehouses and threads.
As shown in Figure~\ref{fig.learned_cc_perf}(b),
%
\dbnamecc adapts quickly to workload drift and outperforms Polyjuice by up to 2.05$\times$. 
The superior performance of \dbnamecc 
mainly stems from its design that encapsulates a fast yet accurate model to find the best concurrency control action, while facilitating the fine-tuning process with the two-phase adaptation algorithm.

\paragraph{Learned Query Optimizer} 
We next compare \dbname with PostgreSQL and two state-of-the-art learned query optimizer approaches, namely Bao~\cite{DBLP:conf/sigmod/MarcusNMTAK21}, and Lero~\cite{DBLP:journals/pvldb/ZhuCDCPWZ23}. 
We construct three workloads with different data distributions and randomly select 8 SPJ queries provided by STATS datasets. 
We use stable models of Bao and Lero for the experiment, as they demonstrated good performance in their respective papers.

\begin{figure}[t]
\centering
\includegraphics[width=0.95\columnwidth]{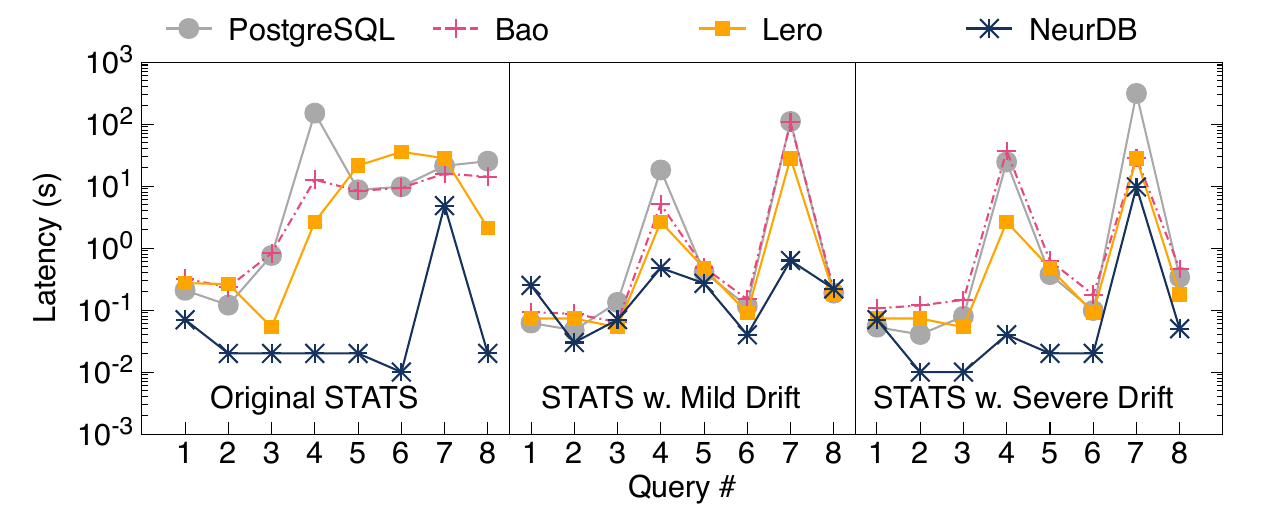}
\begin{minipage}{\columnwidth}
\footnotesize
\end{minipage}
\captionsetup{justification=raggedright}
\caption{Performance of Learned Query Optimizers}
\label{fig.learned_opt_perf}
\vspace{-2mm}
\end{figure}

\noindent
As can be observed in Figure~\ref{fig.learned_opt_perf}, \dbname achieves up to 20.32\% lower average latency of all evaluated queries, which demonstrates its effective adaptability to both data and workload drift.
Due to the 
proposed
dual-module model, \dbname is able to effectively capture system conditions and use them to select an efficient query plan.

\section{Related Work} \label{sec:related}

Our work relates to a broad spectrum of efforts on the fusion of AI and databases.
The initial concepts~\cite{DBLP:conf/aaai/Brodie88, DBLP:conf/sigmod/GraefeD87} can be traced back to the 1980s.
At that time, the fusion was far from reality due to limited development in both realms.
With advancements in AI and database fields over the years, numerous attempts by academia~\cite{DBLP:conf/cidr/KraskaABCKLMMN19,DBLP:conf/cidr/PavloAALLMMMPQS17} and industry~\cite{DBLP:conf/cidr/KaranasosIPSPPX20,DBLP:journals/pvldb/0001ZSYHJLWL21} have been made to advance in-database AI analytics and autonomous DBMS optimization.
However, 
many problems remain open, and it is important to consider data and workload drift in both directions.
Here we shall briefly discuss recent advancements in AI and DBMS integration,  with a particular focus on their approaches to handling data and workload drift.

\revision{
\paragraph{In-database AI Analytics}
Recent research increasingly focuses on optimizing the performance of in-database AI analytics.
For example, some studies~\cite{DBLP:journals/pvldb/SalazarDiazGR24,DBLP:conf/sigmod/ParkSBSIK22,DBLP:conf/cidr/KaranasosIPSPPX20,DBLP:conf/sigmod/Jasny0KRB20} extend the existing query and transaction execution framework to improve the efficiency of analytics tasks.
Further, in-database model management systems~\cite{DBLP:journals/pvldb/SuAWAC24,DBLP:conf/cidr/Vartak17,DBLP:journals/pvldb/ZhouCDMYZZ22} are proposed to minimize model storage costs while efficiently serving models for analytics tasks.
These optimizations can be broadly adopted by full-fledged DBMSs supporting AI analytics, such as nsDB~\cite{DBLP:journals/pvldb/YuanTZZQ24}, GaussML~\cite{DBLP:conf/icde/LiSXLWN24}, and \dbname.
However, apart from the mentioned approaches, \dbname establishes an in-database AI ecosystem, specifically designed to enhance the efficiency and effectiveness of AI analytics under continuous data and workload drift.

\paragraph{Autonomous DBMS Optimization}
Traditional AI-driven DBMS optimization in areas such as knob tuning~\cite{DBLP:conf/cidr/PavloAALLMMMPQS17}, resource management~\cite{DBLP:journals/pvldb/LiZZLZXXFZ023,DBLP:journals/pvldb/0001ZSYHJLWL21}, 
and learned system components~\cite{DBLP:conf/cidr/KraskaABCKLMMN19,DBLP:conf/sigmod/MarcusNMTAK21,DBLP:journals/pvldb/ZhuCDCPWZ23,DBLP:conf/osdi/WangDWCW0021,DBLP:conf/sigmod/KraskaBCDP18} typically relies on complete retraining to handle data and workload drift.
Recently, the development of drift-handling mechanisms become popular, particularly for learned query optimizers and learned indexes.
For example, some works introduce methods to automatically detect drift and trigger model updates~\cite{DBLP:journals/pacmmod/WuI24,DBLP:conf/sigmod/LiLK22,DBLP:journals/pacmmod/KurmanjiT23}, enabling learned query optimizers to withstand data and workload drift.
Further, updatable learned indexes~\cite{DBLP:journals/pacmmod/ZhangQYB24,DBLP:conf/sigmod/DingMYWDLZCGKLK20,DBLP:journals/pvldb/WuZCCWX21} are enhanced with the adaptability to continuous transactional updates.
Unlike existing works focusing on certain components, \dbname introduces a unified architecture based on the proposed in-database AI ecosystem to facilitate fast adaptation across all its key components.


We are extending \dbname to be an intelligent cloud-native HTAP database with built-in support for advanced AI analytics. 
In particular, we disaggregate the system architecture into separate compute and storage layers~\cite{DBLP:journals/pacmmod/Pang024,DBLP:conf/sigmod/CaoZYLWHCCLFWWS21}, enabling dedicated compute nodes to handle OLTP, OLAP, and AI analytics independently.
Based on this foundation, we are devising AI-powered resource scheduling to enable on-demand resource allocation and component assembling, thus enhancing system scalability and adaptability.
}

\section{Conclusions} \label{sec:conclusion}

This paper presents \dbname, a novel AI-powered autonomous DBMS that is adaptable
to data and workload drift.
\dbname,
with its fast-adaptive learned database components and
in-database AI ecosystem,
facilitates
efficient and effective in-database AI analytics and autonomous system optimization.
Empirical evaluations demonstrate the superiority of \dbname, highlighting its potential to realize a seamless fusion of AI and databases.


\begin{acks}
We would like to thank Jiaqi Zhu for her help and comments.
The work of NUS researchers is partially supported by the Lee Foundation in terms of Beng Chin Ooi's Lee Kong Chian Centennial Professorship fund and NUS Faculty Development Fund.
Gang Chen's work is supported by National Key Research and Development Program of China (2022YFB2703100).
Yanyan Shen's work is supported by National Key Research and Development Program of China (2022YFE0200500).
Yuncheng Wu's work is supported by National Key Research and Development Program of China (2023YFB4503600).
Meihui Zhang's work is supported by National Natural Science Foundation of China (62072033).
\end{acks}

\newpage

\bibliographystyle{ACM-Reference-Format}
\bibliography{main-bibliography}


\begin{thebibliography}{55}


\ifx \showCODEN    \undefined \def \showCODEN     #1{\unskip}     \fi
\ifx \showDOI      \undefined \def \showDOI       #1{#1}\fi
\ifx \showISBNx    \undefined \def \showISBNx     #1{\unskip}     \fi
\ifx \showISBNxiii \undefined \def \showISBNxiii  #1{\unskip}     \fi
\ifx \showISSN     \undefined \def \showISSN      #1{\unskip}     \fi
\ifx \showLCCN     \undefined \def \showLCCN      #1{\unskip}     \fi
\ifx \shownote     \undefined \def \shownote      #1{#1}          \fi
\ifx \showarticletitle \undefined \def \showarticletitle #1{#1}   \fi
\ifx \showURL      \undefined \def \showURL       {\relax}        \fi
\providecommand\bibfield[2]{#2}
\providecommand\bibinfo[2]{#2}
\providecommand\natexlab[1]{#1}
\providecommand\showeprint[2][]{arXiv:#2}

\bibitem[{B}ig{Q}uery(2024)]%
        {bigquery}
\bibfield{author}{\bibinfo{person}{Google {B}ig{Q}uery}.} \bibinfo{year}{2024}\natexlab{}.
\newblock \bibinfo{howpublished}{\url{https://cloud.google.com/bigquery/docs}}.
\newblock


\bibitem[Brodie(1988)]%
        {DBLP:conf/aaai/Brodie88}
\bibfield{author}{\bibinfo{person}{Michael~L. Brodie}.} \bibinfo{year}{1988}\natexlab{}.
\newblock \showarticletitle{Future Intelligent Information Systems: {AI} and Database Technologies Working Together}. In \bibinfo{booktitle}{\emph{{AAAI}}}. \bibinfo{publisher}{{AAAI} Press / The {MIT} Press}, \bibinfo{pages}{844--845}.
\newblock


\bibitem[Cai et~al\mbox{.}(2021)]%
        {cai2021arm}
\bibfield{author}{\bibinfo{person}{Shaofeng Cai}, \bibinfo{person}{Kaiping Zheng}, \bibinfo{person}{Gang Chen}, \bibinfo{person}{H.~V. Jagadish}, \bibinfo{person}{Beng~Chin Ooi}, {and} \bibinfo{person}{Meihui Zhang}.} \bibinfo{year}{2021}\natexlab{}.
\newblock \showarticletitle{{ARM}-{N}et: Adaptive Relation Modeling Network for Structured Data}. In \bibinfo{booktitle}{\emph{{SIGMOD} Conference}}. \bibinfo{publisher}{{ACM}}, \bibinfo{pages}{207--220}.
\newblock


\bibitem[Cao et~al\mbox{.}(2021)]%
        {DBLP:conf/sigmod/CaoZYLWHCCLFWWS21}
\bibfield{author}{\bibinfo{person}{Wei Cao}, \bibinfo{person}{Yingqiang Zhang}, \bibinfo{person}{Xinjun Yang}, \bibinfo{person}{Feifei Li}, \bibinfo{person}{Sheng Wang}, \bibinfo{person}{Qingda Hu}, \bibinfo{person}{Xuntao Cheng}, \bibinfo{person}{Zongzhi Chen}, \bibinfo{person}{Zhenjun Liu}, \bibinfo{person}{Jing Fang}, \bibinfo{person}{Bo Wang}, \bibinfo{person}{Yuhui Wang}, \bibinfo{person}{Haiqing Sun}, \bibinfo{person}{Ze Yang}, \bibinfo{person}{Zhushi Cheng}, \bibinfo{person}{Sen Chen}, \bibinfo{person}{Jian Wu}, \bibinfo{person}{Wei Hu}, \bibinfo{person}{Jianwei Zhao}, \bibinfo{person}{Yusong Gao}, \bibinfo{person}{Songlu Cai}, \bibinfo{person}{Yunyang Zhang}, {and} \bibinfo{person}{Jiawang Tong}.} \bibinfo{year}{2021}\natexlab{}.
\newblock \showarticletitle{PolarDB Serverless: {A} Cloud Native Database for Disaggregated Data Centers}. In \bibinfo{booktitle}{\emph{{SIGMOD} Conference}}. \bibinfo{publisher}{{ACM}}, \bibinfo{pages}{2477--2489}.
\newblock


\bibitem[Cong et~al\mbox{.}(2024)]%
        {DBLP:conf/sigmod/CongYZ24}
\bibfield{author}{\bibinfo{person}{Gao Cong}, \bibinfo{person}{Jingyi Yang}, {and} \bibinfo{person}{Yue Zhao}.} \bibinfo{year}{2024}\natexlab{}.
\newblock \showarticletitle{Machine Learning for Databases: Foundations, Paradigms, and Open problems}. In \bibinfo{booktitle}{\emph{{SIGMOD} Conference Companion}}. \bibinfo{publisher}{{ACM}}, \bibinfo{pages}{622--629}.
\newblock


\bibitem[Cooper et~al\mbox{.}(2010)]%
        {DBLP:conf/cloud/CooperSTRS10}
\bibfield{author}{\bibinfo{person}{Brian~F. Cooper}, \bibinfo{person}{Adam Silberstein}, \bibinfo{person}{Erwin Tam}, \bibinfo{person}{Raghu Ramakrishnan}, {and} \bibinfo{person}{Russell Sears}.} \bibinfo{year}{2010}\natexlab{}.
\newblock \showarticletitle{Benchmarking cloud serving systems with {YCSB}}. In \bibinfo{booktitle}{\emph{SoCC}}. \bibinfo{publisher}{{ACM}}, \bibinfo{pages}{143--154}.
\newblock


\bibitem[Dataset(2024a)]%
        {avazu}
\bibfield{author}{\bibinfo{person}{Avazu Dataset}.} \bibinfo{year}{2024}\natexlab{a}.
\newblock \bibinfo{howpublished}{\url{https://www.kaggle.com/c/avazu-ctr-prediction}}.
\newblock


\bibitem[Dataset(2024b)]%
        {stats}
\bibfield{author}{\bibinfo{person}{{STATS} Dataset}.} \bibinfo{year}{2024}\natexlab{b}.
\newblock \bibinfo{howpublished}{\url{https://github.com/Nathaniel-Han/End-to-End-CardEst-Benchmark}}.
\newblock


\bibitem[Dataset(2024c)]%
        {diabetes}
\bibfield{author}{\bibinfo{person}{{UCI}~Diabetes Dataset}.} \bibinfo{year}{2024}\natexlab{c}.
\newblock \bibinfo{howpublished}{\url{https://archive.ics.uci.edu/ml/datasets}}.
\newblock


\bibitem[Ding et~al\mbox{.}(2020)]%
        {DBLP:conf/sigmod/DingMYWDLZCGKLK20}
\bibfield{author}{\bibinfo{person}{Jialin Ding}, \bibinfo{person}{Umar~Farooq Minhas}, \bibinfo{person}{Jia Yu}, \bibinfo{person}{Chi Wang}, \bibinfo{person}{Jaeyoung Do}, \bibinfo{person}{Yinan Li}, \bibinfo{person}{Hantian Zhang}, \bibinfo{person}{Badrish Chandramouli}, \bibinfo{person}{Johannes Gehrke}, \bibinfo{person}{Donald Kossmann}, \bibinfo{person}{David~B. Lomet}, {and} \bibinfo{person}{Tim Kraska}.} \bibinfo{year}{2020}\natexlab{}.
\newblock \showarticletitle{{ALEX:} An Updatable Adaptive Learned Index}. In \bibinfo{booktitle}{\emph{{SIGMOD} Conference}}. \bibinfo{publisher}{{ACM}}, \bibinfo{pages}{969--984}.
\newblock


\bibitem[Filter and Refine(2024)]%
        {frp}
\bibfield{author}{\bibinfo{person}{Filter} {and} \bibinfo{person}{Refine}.} \bibinfo{year}{2024}\natexlab{}.
\newblock \bibinfo{howpublished}{\url{https://en.wikipedia.org/wiki/Filter_and_refine/}}.
\newblock


\bibitem[Graefe and DeWitt(1987)]%
        {DBLP:conf/sigmod/GraefeD87}
\bibfield{author}{\bibinfo{person}{Goetz Graefe} {and} \bibinfo{person}{David~J. DeWitt}.} \bibinfo{year}{1987}\natexlab{}.
\newblock \showarticletitle{The {EXODUS} Optimizer Generator}. In \bibinfo{booktitle}{\emph{{SIGMOD} Conference}}. \bibinfo{publisher}{{ACM} Press}, \bibinfo{pages}{160--172}.
\newblock


\bibitem[Jasny et~al\mbox{.}(2020)]%
        {DBLP:conf/sigmod/Jasny0KRB20}
\bibfield{author}{\bibinfo{person}{Matthias Jasny}, \bibinfo{person}{Tobias Ziegler}, \bibinfo{person}{Tim Kraska}, \bibinfo{person}{Uwe R{\"{o}}hm}, {and} \bibinfo{person}{Carsten Binnig}.} \bibinfo{year}{2020}\natexlab{}.
\newblock \showarticletitle{{DB4ML} - An In-Memory Database Kernel with Machine Learning Support}. In \bibinfo{booktitle}{\emph{{SIGMOD} Conference}}. \bibinfo{publisher}{{ACM}}, \bibinfo{pages}{159--173}.
\newblock


\bibitem[Karanasos et~al\mbox{.}(2020)]%
        {DBLP:conf/cidr/KaranasosIPSPPX20}
\bibfield{author}{\bibinfo{person}{Konstantinos Karanasos}, \bibinfo{person}{Matteo Interlandi}, \bibinfo{person}{Fotis Psallidas}, \bibinfo{person}{Rathijit Sen}, \bibinfo{person}{Kwanghyun Park}, \bibinfo{person}{Ivan Popivanov}, \bibinfo{person}{Doris Xin}, \bibinfo{person}{Supun Nakandala}, \bibinfo{person}{Subru Krishnan}, \bibinfo{person}{Markus Weimer}, \bibinfo{person}{Yuan Yu}, \bibinfo{person}{Raghu Ramakrishnan}, {and} \bibinfo{person}{Carlo Curino}.} \bibinfo{year}{2020}\natexlab{}.
\newblock \showarticletitle{Extending Relational Query Processing with {ML} Inference}. In \bibinfo{booktitle}{\emph{{CIDR}}}. \bibinfo{publisher}{www.cidrdb.org}.
\newblock


\bibitem[Kraska et~al\mbox{.}(2019)]%
        {DBLP:conf/cidr/KraskaABCKLMMN19}
\bibfield{author}{\bibinfo{person}{Tim Kraska}, \bibinfo{person}{Mohammad Alizadeh}, \bibinfo{person}{Alex Beutel}, \bibinfo{person}{Ed~H. Chi}, \bibinfo{person}{Ani Kristo}, \bibinfo{person}{Guillaume Leclerc}, \bibinfo{person}{Samuel Madden}, \bibinfo{person}{Hongzi Mao}, {and} \bibinfo{person}{Vikram Nathan}.} \bibinfo{year}{2019}\natexlab{}.
\newblock \showarticletitle{SageDB: {A} Learned Database System}. In \bibinfo{booktitle}{\emph{{CIDR}}}. \bibinfo{publisher}{www.cidrdb.org}.
\newblock


\bibitem[Kraska et~al\mbox{.}(2018)]%
        {DBLP:conf/sigmod/KraskaBCDP18}
\bibfield{author}{\bibinfo{person}{Tim Kraska}, \bibinfo{person}{Alex Beutel}, \bibinfo{person}{Ed~H. Chi}, \bibinfo{person}{Jeffrey Dean}, {and} \bibinfo{person}{Neoklis Polyzotis}.} \bibinfo{year}{2018}\natexlab{}.
\newblock \showarticletitle{The Case for Learned Index Structures}. In \bibinfo{booktitle}{\emph{{SIGMOD} Conference}}. \bibinfo{publisher}{{ACM}}, \bibinfo{pages}{489--504}.
\newblock


\bibitem[Kurmanji et~al\mbox{.}(2024)]%
        {DBLP:journals/pacmmod/KurmanjiTT24}
\bibfield{author}{\bibinfo{person}{Meghdad Kurmanji}, \bibinfo{person}{Eleni Triantafillou}, {and} \bibinfo{person}{Peter Triantafillou}.} \bibinfo{year}{2024}\natexlab{}.
\newblock \showarticletitle{Machine Unlearning in Learned Databases: An Experimental Analysis}.
\newblock \bibinfo{journal}{\emph{Proc. {ACM} Manag. Data}} \bibinfo{volume}{2}, \bibinfo{number}{1} (\bibinfo{year}{2024}), \bibinfo{pages}{49:1--49:26}.
\newblock


\bibitem[Kurmanji and Triantafillou(2023)]%
        {DBLP:journals/pacmmod/KurmanjiT23}
\bibfield{author}{\bibinfo{person}{Meghdad Kurmanji} {and} \bibinfo{person}{Peter Triantafillou}.} \bibinfo{year}{2023}\natexlab{}.
\newblock \showarticletitle{Detect, Distill and Update: Learned {DB} Systems Facing Out of Distribution Data}.
\newblock \bibinfo{journal}{\emph{Proc. {ACM} Manag. Data}} \bibinfo{volume}{1}, \bibinfo{number}{1} (\bibinfo{year}{2023}), \bibinfo{pages}{33:1--33:27}.
\newblock


\bibitem[Li et~al\mbox{.}(2022)]%
        {DBLP:conf/sigmod/LiLK22}
\bibfield{author}{\bibinfo{person}{Beibin Li}, \bibinfo{person}{Yao Lu}, {and} \bibinfo{person}{Srikanth Kandula}.} \bibinfo{year}{2022}\natexlab{}.
\newblock \showarticletitle{Warper: Efficiently Adapting Learned Cardinality Estimators to Data and Workload Drifts}. In \bibinfo{booktitle}{\emph{{SIGMOD} Conference}}. \bibinfo{publisher}{{ACM}}, \bibinfo{pages}{1920--1933}.
\newblock


\bibitem[Li et~al\mbox{.}(2024)]%
        {DBLP:conf/icde/LiSXLWN24}
\bibfield{author}{\bibinfo{person}{Guoliang Li}, \bibinfo{person}{Ji Sun}, \bibinfo{person}{Lijie Xu}, \bibinfo{person}{Shifu Li}, \bibinfo{person}{Jiang Wang}, {and} \bibinfo{person}{Wen Nie}.} \bibinfo{year}{2024}\natexlab{}.
\newblock \showarticletitle{GaussML: An End-to-End In-Database Machine Learning System}. In \bibinfo{booktitle}{\emph{{ICDE}}}. \bibinfo{publisher}{{IEEE}}, \bibinfo{pages}{5198--5210}.
\newblock


\bibitem[Li et~al\mbox{.}(2021)]%
        {DBLP:journals/pvldb/0001ZSYHJLWL21}
\bibfield{author}{\bibinfo{person}{Guoliang Li}, \bibinfo{person}{Xuanhe Zhou}, \bibinfo{person}{Ji Sun}, \bibinfo{person}{Xiang Yu}, \bibinfo{person}{Yue Han}, \bibinfo{person}{Lianyuan Jin}, \bibinfo{person}{Wenbo Li}, \bibinfo{person}{Tianqing Wang}, {and} \bibinfo{person}{Shifu Li}.} \bibinfo{year}{2021}\natexlab{}.
\newblock \showarticletitle{openGauss: An Autonomous Database System}.
\newblock \bibinfo{journal}{\emph{Proc. {VLDB} Endow.}} \bibinfo{volume}{14}, \bibinfo{number}{12} (\bibinfo{year}{2021}), \bibinfo{pages}{3028--3041}.
\newblock


\bibitem[Li et~al\mbox{.}(2023b)]%
        {DBLP:journals/pvldb/LiZZLZXXFZ023}
\bibfield{author}{\bibinfo{person}{Ji{-}You Li}, \bibinfo{person}{Jiachi Zhang}, \bibinfo{person}{Wenchao Zhou}, \bibinfo{person}{Yuhang Liu}, \bibinfo{person}{Shuai Zhang}, \bibinfo{person}{Zhuoming Xue}, \bibinfo{person}{Ding Xu}, \bibinfo{person}{Hua Fan}, \bibinfo{person}{Fangyuan Zhou}, {and} \bibinfo{person}{Feifei Li}.} \bibinfo{year}{2023}\natexlab{b}.
\newblock \showarticletitle{Eigen: End-to-end Resource Optimization for Large-Scale Databases on the Cloud}.
\newblock \bibinfo{journal}{\emph{Proc. {VLDB} Endow.}} \bibinfo{volume}{16}, \bibinfo{number}{12} (\bibinfo{year}{2023}), \bibinfo{pages}{3795--3807}.
\newblock


\bibitem[Li et~al\mbox{.}(2023a)]%
        {DBLP:journals/pvldb/LiWZDZ023}
\bibfield{author}{\bibinfo{person}{Pengfei Li}, \bibinfo{person}{Wenqing Wei}, \bibinfo{person}{Rong Zhu}, \bibinfo{person}{Bolin Ding}, \bibinfo{person}{Jingren Zhou}, {and} \bibinfo{person}{Hua Lu}.} \bibinfo{year}{2023}\natexlab{a}.
\newblock \showarticletitle{{ALECE:} An Attention-based Learned Cardinality Estimator for {SPJ} Queries on Dynamic Workloads}.
\newblock \bibinfo{journal}{\emph{Proc. {VLDB} Endow.}} \bibinfo{volume}{17}, \bibinfo{number}{2} (\bibinfo{year}{2023}), \bibinfo{pages}{197--210}.
\newblock


\bibitem[Marcus et~al\mbox{.}(2021)]%
        {DBLP:conf/sigmod/MarcusNMTAK21}
\bibfield{author}{\bibinfo{person}{Ryan Marcus}, \bibinfo{person}{Parimarjan Negi}, \bibinfo{person}{Hongzi Mao}, \bibinfo{person}{Nesime Tatbul}, \bibinfo{person}{Mohammad Alizadeh}, {and} \bibinfo{person}{Tim Kraska}.} \bibinfo{year}{2021}\natexlab{}.
\newblock \showarticletitle{Bao: Making Learned Query Optimization Practical}. In \bibinfo{booktitle}{\emph{{SIGMOD} Conference}}. \bibinfo{publisher}{{ACM}}, \bibinfo{pages}{1275--1288}.
\newblock


\bibitem[Marcus et~al\mbox{.}(2019)]%
        {DBLP:journals/pvldb/MarcusNMZAKPT19}
\bibfield{author}{\bibinfo{person}{Ryan Marcus}, \bibinfo{person}{Parimarjan Negi}, \bibinfo{person}{Hongzi Mao}, \bibinfo{person}{Chi Zhang}, \bibinfo{person}{Mohammad Alizadeh}, \bibinfo{person}{Tim Kraska}, \bibinfo{person}{Olga Papaemmanouil}, {and} \bibinfo{person}{Nesime Tatbul}.} \bibinfo{year}{2019}\natexlab{}.
\newblock \showarticletitle{Neo: {A} Learned Query Optimizer}.
\newblock \bibinfo{journal}{\emph{Proc. {VLDB} Endow.}} \bibinfo{volume}{12}, \bibinfo{number}{11} (\bibinfo{year}{2019}), \bibinfo{pages}{1705--1718}.
\newblock


\bibitem[Minds{DB}(2024)]%
        {mindsdb}
\bibfield{author}{\bibinfo{person}{Minds{DB}}.} \bibinfo{year}{2024}\natexlab{}.
\newblock \bibinfo{howpublished}{\url{https://mindsdb.com/}}.
\newblock


\bibitem[NeurDB(2024)]%
        {neurdbcode}
\bibfield{author}{\bibinfo{person}{NeurDB}.} \bibinfo{year}{2024}\natexlab{}.
\newblock \bibinfo{howpublished}{\url{https://github.com/neurdb/neurdb}}.
\newblock


\bibitem[Ooi et~al\mbox{.}(2024)]%
        {neurdb-scis-24}
\bibfield{author}{\bibinfo{person}{Beng~Chin Ooi}, \bibinfo{person}{Shaofeng Cai}, \bibinfo{person}{Gang Chen}, \bibinfo{person}{Yanyan Shen}, \bibinfo{person}{Kian-Lee Tan}, \bibinfo{person}{Yuncheng Wu}, \bibinfo{person}{Xiaokui Xiao}, \bibinfo{person}{Naili Xing}, \bibinfo{person}{Cong Yue}, \bibinfo{person}{Lingze Zeng}, \bibinfo{person}{Meihui Zhang}, {and} \bibinfo{person}{Zhanhao Zhao}.} \bibinfo{year}{2024}\natexlab{}.
\newblock \showarticletitle{NeurDB: An AI-powered Autonomous Data System}.
\newblock \bibinfo{journal}{\emph{SCIENCE CHINA Information Sciences}} \bibinfo{volume}{67}, \bibinfo{number}{200901} (\bibinfo{year}{2024}).
\newblock
\urldef\tempurl%
\url{https://doi.org/10.1007/s11432-024-4125-9}
\showDOI{\tempurl}


\bibitem[Ooi et~al\mbox{.}(2002)]%
        {DBLP:conf/ideas/OoiPWWY02}
\bibfield{author}{\bibinfo{person}{Beng~Chin Ooi}, \bibinfo{person}{HweeHwa Pang}, \bibinfo{person}{Hao Wang}, \bibinfo{person}{Limsoon Wong}, {and} \bibinfo{person}{Cui Yu}.} \bibinfo{year}{2002}\natexlab{}.
\newblock \showarticletitle{Fast Filter-and-Refine Algorithms for Subsequence Selection}. In \bibinfo{booktitle}{\emph{{IDEAS}}}. \bibinfo{publisher}{{IEEE} Computer Society}, \bibinfo{pages}{243--255}.
\newblock


\bibitem[Ooi et~al\mbox{.}(2015)]%
        {DBLP:conf/mm/OoiTWWCCGLTWXZZ15}
\bibfield{author}{\bibinfo{person}{Beng~Chin Ooi}, \bibinfo{person}{Kian{-}Lee Tan}, \bibinfo{person}{Sheng Wang}, \bibinfo{person}{Wei Wang}, \bibinfo{person}{Qingchao Cai}, \bibinfo{person}{Gang Chen}, \bibinfo{person}{Jinyang Gao}, \bibinfo{person}{Zhaojing Luo}, \bibinfo{person}{Anthony K.~H. Tung}, \bibinfo{person}{Yuan Wang}, \bibinfo{person}{Zhongle Xie}, \bibinfo{person}{Meihui Zhang}, {and} \bibinfo{person}{Kaiping Zheng}.} \bibinfo{year}{2015}\natexlab{}.
\newblock \showarticletitle{{SINGA:} {A} Distributed Deep Learning Platform}. In \bibinfo{booktitle}{\emph{{ACM} Multimedia}}. \bibinfo{publisher}{{ACM}}, \bibinfo{pages}{685--688}.
\newblock


\bibitem[Oracle(2024)]%
        {oracle}
\bibfield{author}{\bibinfo{person}{Oracle}.} \bibinfo{year}{2024}\natexlab{}.
\newblock \bibinfo{howpublished}{\url{https://docs.oracle.com/en/database/oracle/machine-learning/}}.
\newblock


\bibitem[Pang and Wang(2024)]%
        {DBLP:journals/pacmmod/Pang024}
\bibfield{author}{\bibinfo{person}{Xi Pang} {and} \bibinfo{person}{Jianguo Wang}.} \bibinfo{year}{2024}\natexlab{}.
\newblock \showarticletitle{Understanding the Performance Implications of the Design Principles in Storage-Disaggregated Databases}.
\newblock \bibinfo{journal}{\emph{Proc. {ACM} Manag. Data}} \bibinfo{volume}{2}, \bibinfo{number}{3} (\bibinfo{year}{2024}), \bibinfo{pages}{180}.
\newblock


\bibitem[Park et~al\mbox{.}(2022)]%
        {DBLP:conf/sigmod/ParkSBSIK22}
\bibfield{author}{\bibinfo{person}{Kwanghyun Park}, \bibinfo{person}{Karla Saur}, \bibinfo{person}{Dalitso Banda}, \bibinfo{person}{Rathijit Sen}, \bibinfo{person}{Matteo Interlandi}, {and} \bibinfo{person}{Konstantinos Karanasos}.} \bibinfo{year}{2022}\natexlab{}.
\newblock \showarticletitle{End-to-end Optimization of Machine Learning Prediction Queries}. In \bibinfo{booktitle}{\emph{{SIGMOD} Conference}}. \bibinfo{publisher}{{ACM}}, \bibinfo{pages}{587--601}.
\newblock


\bibitem[Pavlo et~al\mbox{.}(2017)]%
        {DBLP:conf/cidr/PavloAALLMMMPQS17}
\bibfield{author}{\bibinfo{person}{Andrew Pavlo}, \bibinfo{person}{Gustavo Angulo}, \bibinfo{person}{Joy Arulraj}, \bibinfo{person}{Haibin Lin}, \bibinfo{person}{Jiexi Lin}, \bibinfo{person}{Lin Ma}, \bibinfo{person}{Prashanth Menon}, \bibinfo{person}{Todd~C. Mowry}, \bibinfo{person}{Matthew Perron}, \bibinfo{person}{Ian Quah}, \bibinfo{person}{Siddharth Santurkar}, \bibinfo{person}{Anthony Tomasic}, \bibinfo{person}{Skye Toor}, \bibinfo{person}{Dana~Van Aken}, \bibinfo{person}{Ziqi Wang}, \bibinfo{person}{Yingjun Wu}, \bibinfo{person}{Ran Xian}, {and} \bibinfo{person}{Tieying Zhang}.} \bibinfo{year}{2017}\natexlab{}.
\newblock \showarticletitle{Self-Driving Database Management Systems}. In \bibinfo{booktitle}{\emph{{CIDR}}}. \bibinfo{publisher}{www.cidrdb.org}.
\newblock


\bibitem[Ports and Grittner(2012)]%
        {DBLP:journals/pvldb/PortsG12}
\bibfield{author}{\bibinfo{person}{Dan R.~K. Ports} {and} \bibinfo{person}{Kevin Grittner}.} \bibinfo{year}{2012}\natexlab{}.
\newblock \showarticletitle{Serializable Snapshot Isolation in PostgreSQL}.
\newblock \bibinfo{journal}{\emph{Proc. {VLDB} Endow.}} \bibinfo{volume}{5}, \bibinfo{number}{12} (\bibinfo{year}{2012}), \bibinfo{pages}{1850--1861}.
\newblock


\bibitem[Postgres{ML}(2024)]%
        {postgresml}
\bibfield{author}{\bibinfo{person}{Postgres{ML}}.} \bibinfo{year}{2024}\natexlab{}.
\newblock \bibinfo{howpublished}{\url{https://postgresml.org/}}.
\newblock


\bibitem[R{\'{e}} et~al\mbox{.}(2015)]%
        {DBLP:conf/sigmod/ReABCJKR15}
\bibfield{author}{\bibinfo{person}{Christopher R{\'{e}}}, \bibinfo{person}{Divy Agrawal}, \bibinfo{person}{Magdalena Balazinska}, \bibinfo{person}{Michael~J. Cafarella}, \bibinfo{person}{Michael~I. Jordan}, \bibinfo{person}{Tim Kraska}, {and} \bibinfo{person}{Raghu Ramakrishnan}.} \bibinfo{year}{2015}\natexlab{}.
\newblock \showarticletitle{Machine Learning and Databases: The Sound of Things to Come or a Cacophony of Hype?}. In \bibinfo{booktitle}{\emph{{SIGMOD} Conference}}. \bibinfo{publisher}{{ACM}}, \bibinfo{pages}{283--284}.
\newblock


\bibitem[{R}edshift(2024)]%
        {amazon}
\bibfield{author}{\bibinfo{person}{Amazon {R}edshift}.} \bibinfo{year}{2024}\natexlab{}.
\newblock \bibinfo{howpublished}{\url{https://docs.aws.amazon.com/redshift/latest/dg/machine_learning.html}}.
\newblock


\bibitem[Salazar{-}D{\'{\i}}az et~al\mbox{.}(2024)]%
        {DBLP:journals/pvldb/SalazarDiazGR24}
\bibfield{author}{\bibinfo{person}{Ricardo Salazar{-}D{\'{\i}}az}, \bibinfo{person}{Boris Glavic}, {and} \bibinfo{person}{Tilmann Rabl}.} \bibinfo{year}{2024}\natexlab{}.
\newblock \showarticletitle{InferDB: In-Database Machine Learning Inference Using Indexes}.
\newblock \bibinfo{journal}{\emph{Proc. {VLDB} Endow.}} \bibinfo{volume}{17}, \bibinfo{number}{8} (\bibinfo{year}{2024}), \bibinfo{pages}{1830--1842}.
\newblock


\bibitem[{SQL}(2024)]%
        {azure}
\bibfield{author}{\bibinfo{person}{Azure {SQL}}.} \bibinfo{year}{2024}\natexlab{}.
\newblock \bibinfo{howpublished}{\url{https://learn.microsoft.com/en-us/azure/azure-sql/}}.
\newblock


\bibitem[Su et~al\mbox{.}(2024)]%
        {DBLP:journals/pvldb/SuAWAC24}
\bibfield{author}{\bibinfo{person}{Zhaoyuan Su}, \bibinfo{person}{Ammar Ahmed}, \bibinfo{person}{Zirui Wang}, \bibinfo{person}{Ali Anwar}, {and} \bibinfo{person}{Yue Cheng}.} \bibinfo{year}{2024}\natexlab{}.
\newblock \showarticletitle{Everything You Always Wanted to Know About Storage Compressibility of Pre-Trained {ML} Models but Were Afraid to Ask}.
\newblock \bibinfo{journal}{\emph{Proc. {VLDB} Endow.}} \bibinfo{volume}{17}, \bibinfo{number}{8} (\bibinfo{year}{2024}), \bibinfo{pages}{2036--2049}.
\newblock


\bibitem[TPCC(2024)]%
        {tpcc}
\bibfield{author}{\bibinfo{person}{TPCC}.} \bibinfo{year}{2024}\natexlab{}.
\newblock \bibinfo{howpublished}{\url{http://www.tpc.org/tpcc/}}.
\newblock


\bibitem[Vartak(2017)]%
        {DBLP:conf/cidr/Vartak17}
\bibfield{author}{\bibinfo{person}{Manasi Vartak}.} \bibinfo{year}{2017}\natexlab{}.
\newblock \showarticletitle{{MODELDB:} {A} System for Machine Learning Model Management}. In \bibinfo{booktitle}{\emph{{CIDR}}}. \bibinfo{publisher}{www.cidrdb.org}.
\newblock


\bibitem[Wang et~al\mbox{.}(2021)]%
        {DBLP:conf/osdi/WangDWCW0021}
\bibfield{author}{\bibinfo{person}{Jia{-}Chen Wang}, \bibinfo{person}{Ding Ding}, \bibinfo{person}{Huan Wang}, \bibinfo{person}{Conrad Christensen}, \bibinfo{person}{Zhaoguo Wang}, \bibinfo{person}{Haibo Chen}, {and} \bibinfo{person}{Jinyang Li}.} \bibinfo{year}{2021}\natexlab{}.
\newblock \showarticletitle{Polyjuice: High-Performance Transactions via Learned Concurrency Control}. In \bibinfo{booktitle}{\emph{{OSDI}}}. \bibinfo{publisher}{{USENIX} Association}, \bibinfo{pages}{198--216}.
\newblock


\bibitem[Wang et~al\mbox{.}(2015)]%
        {DBLP:conf/mm/WangCDGOTW15}
\bibfield{author}{\bibinfo{person}{Wei Wang}, \bibinfo{person}{Gang Chen}, \bibinfo{person}{Tien Tuan~Anh Dinh}, \bibinfo{person}{Jinyang Gao}, \bibinfo{person}{Beng~Chin Ooi}, \bibinfo{person}{Kian{-}Lee Tan}, {and} \bibinfo{person}{Sheng Wang}.} \bibinfo{year}{2015}\natexlab{}.
\newblock \showarticletitle{{SINGA:} Putting Deep Learning in the Hands of Multimedia Users}. In \bibinfo{booktitle}{\emph{{ACM} Multimedia}}. \bibinfo{publisher}{{ACM}}, \bibinfo{pages}{25--34}.
\newblock


\bibitem[Wang et~al\mbox{.}(2016)]%
        {DBLP:journals/sigmod/0059Z0JOT16}
\bibfield{author}{\bibinfo{person}{Wei Wang}, \bibinfo{person}{Meihui Zhang}, \bibinfo{person}{Gang Chen}, \bibinfo{person}{H.~V. Jagadish}, \bibinfo{person}{Beng~Chin Ooi}, {and} \bibinfo{person}{Kian{-}Lee Tan}.} \bibinfo{year}{2016}\natexlab{}.
\newblock \showarticletitle{Database Meets Deep Learning: Challenges and Opportunities}.
\newblock \bibinfo{journal}{\emph{{SIGMOD} Rec.}} \bibinfo{volume}{45}, \bibinfo{number}{2} (\bibinfo{year}{2016}), \bibinfo{pages}{17--22}.
\newblock


\bibitem[Wu et~al\mbox{.}(2021)]%
        {DBLP:journals/pvldb/WuZCCWX21}
\bibfield{author}{\bibinfo{person}{Jiacheng Wu}, \bibinfo{person}{Yong Zhang}, \bibinfo{person}{Shimin Chen}, \bibinfo{person}{Yu Chen}, \bibinfo{person}{Jin Wang}, {and} \bibinfo{person}{Chunxiao Xing}.} \bibinfo{year}{2021}\natexlab{}.
\newblock \showarticletitle{Updatable Learned Index with Precise Positions}.
\newblock \bibinfo{journal}{\emph{Proc. {VLDB} Endow.}} \bibinfo{volume}{14}, \bibinfo{number}{8} (\bibinfo{year}{2021}), \bibinfo{pages}{1276--1288}.
\newblock


\bibitem[Wu and Ives(2024)]%
        {DBLP:journals/pacmmod/WuI24}
\bibfield{author}{\bibinfo{person}{Peizhi Wu} {and} \bibinfo{person}{Zachary~G. Ives}.} \bibinfo{year}{2024}\natexlab{}.
\newblock \showarticletitle{Modeling Shifting Workloads for Learned Database Systems}.
\newblock \bibinfo{journal}{\emph{Proc. {ACM} Manag. Data}} \bibinfo{volume}{2}, \bibinfo{number}{1} (\bibinfo{year}{2024}), \bibinfo{pages}{38:1--38:27}.
\newblock


\bibitem[Xing et~al\mbox{.}(2024)]%
        {DBLP:journals/pvldb/XingCCLOP24}
\bibfield{author}{\bibinfo{person}{Naili Xing}, \bibinfo{person}{Shaofeng Cai}, \bibinfo{person}{Gang Chen}, \bibinfo{person}{Zhaojing Luo}, \bibinfo{person}{Beng~Chin Ooi}, {and} \bibinfo{person}{Jian Pei}.} \bibinfo{year}{2024}\natexlab{}.
\newblock \showarticletitle{Database Native Model Selection: Harnessing Deep Neural Networks in Database Systems}.
\newblock \bibinfo{journal}{\emph{Proc. {VLDB} Endow.}} \bibinfo{volume}{17}, \bibinfo{number}{5} (\bibinfo{year}{2024}), \bibinfo{pages}{1020--1033}.
\newblock


\bibitem[Yuan et~al\mbox{.}(2024)]%
        {DBLP:journals/pvldb/YuanTZZQ24}
\bibfield{author}{\bibinfo{person}{Ye Yuan}, \bibinfo{person}{Bo Tang}, \bibinfo{person}{Tianfei Zhou}, \bibinfo{person}{Zhiwei Zhang}, {and} \bibinfo{person}{Jianbin Qin}.} \bibinfo{year}{2024}\natexlab{}.
\newblock \showarticletitle{nsDB: Architecting the Next Generation Database by Integrating Neural and Symbolic Systems (Vision)}.
\newblock \bibinfo{journal}{\emph{Proc. {VLDB} Endow.}} \bibinfo{volume}{17}, \bibinfo{number}{11} (\bibinfo{year}{2024}), \bibinfo{pages}{3283--3289}.
\newblock


\bibitem[Zeng et~al\mbox{.}(2025)]%
        {DBLP:journals/pvldb/zengslicing25}
\bibfield{author}{\bibinfo{person}{Lingze Zeng}, \bibinfo{person}{Naili Xing}, \bibinfo{person}{Shaofeng Cai}, \bibinfo{person}{Gang Chen}, \bibinfo{person}{Beng~Chin Ooi}, \bibinfo{person}{Jian Pei}, {and} \bibinfo{person}{Yuncheng Wu}.} \bibinfo{year}{2025}\natexlab{}.
\newblock \showarticletitle{Powering In-Database Dynamic Model Slicing for Structured Data Analytics}.
\newblock \bibinfo{journal}{\emph{Proc. {VLDB} Endow.}} \bibinfo{volume}{17}, \bibinfo{number}{1} (\bibinfo{year}{2025}), \bibinfo{pages}{1020--1033}.
\newblock


\bibitem[Zhang et~al\mbox{.}(2024)]%
        {DBLP:journals/pacmmod/ZhangQYB24}
\bibfield{author}{\bibinfo{person}{Shunkang Zhang}, \bibinfo{person}{Ji Qi}, \bibinfo{person}{Xin Yao}, {and} \bibinfo{person}{Andr{\'{e}} Brinkmann}.} \bibinfo{year}{2024}\natexlab{}.
\newblock \showarticletitle{Hyper: {A} High-Performance and Memory-Efficient Learned Index via Hybrid Construction}.
\newblock \bibinfo{journal}{\emph{Proc. {ACM} Manag. Data}} \bibinfo{volume}{2}, \bibinfo{number}{3} (\bibinfo{year}{2024}), \bibinfo{pages}{145}.
\newblock


\bibitem[Zhou et~al\mbox{.}(2022)]%
        {DBLP:journals/pvldb/ZhouCDMYZZ22}
\bibfield{author}{\bibinfo{person}{Lixi Zhou}, \bibinfo{person}{Jiaqing Chen}, \bibinfo{person}{Amitabh Das}, \bibinfo{person}{Hong Min}, \bibinfo{person}{Lei Yu}, \bibinfo{person}{Ming Zhao}, {and} \bibinfo{person}{Jia Zou}.} \bibinfo{year}{2022}\natexlab{}.
\newblock \showarticletitle{Serving Deep Learning Models with Deduplication from Relational Databases}.
\newblock \bibinfo{journal}{\emph{Proc. {VLDB} Endow.}} \bibinfo{volume}{15}, \bibinfo{number}{10} (\bibinfo{year}{2022}), \bibinfo{pages}{2230--2243}.
\newblock


\bibitem[Zhu et~al\mbox{.}(2023)]%
        {DBLP:journals/pvldb/ZhuCDCPWZ23}
\bibfield{author}{\bibinfo{person}{Rong Zhu}, \bibinfo{person}{Wei Chen}, \bibinfo{person}{Bolin Ding}, \bibinfo{person}{Xingguang Chen}, \bibinfo{person}{Andreas Pfadler}, \bibinfo{person}{Ziniu Wu}, {and} \bibinfo{person}{Jingren Zhou}.} \bibinfo{year}{2023}\natexlab{}.
\newblock \showarticletitle{Lero: {A} Learning-to-Rank Query Optimizer}.
\newblock \bibinfo{journal}{\emph{Proc. {VLDB} Endow.}} \bibinfo{volume}{16}, \bibinfo{number}{6} (\bibinfo{year}{2023}), \bibinfo{pages}{1466--1479}.
\newblock


\bibitem[Zhu et~al\mbox{.}(2024)]%
        {DBLP:journals/pvldb/ZhuWWWPWDLZZ24}
\bibfield{author}{\bibinfo{person}{Rong Zhu}, \bibinfo{person}{Lianggui Weng}, \bibinfo{person}{Wenqing Wei}, \bibinfo{person}{Di Wu}, \bibinfo{person}{Jiazhen Peng}, \bibinfo{person}{Yifan Wang}, \bibinfo{person}{Bolin Ding}, \bibinfo{person}{Defu Lian}, \bibinfo{person}{Bolong Zheng}, {and} \bibinfo{person}{Jingren Zhou}.} \bibinfo{year}{2024}\natexlab{}.
\newblock \showarticletitle{PilotScope: Steering Databases with Machine Learning Drivers}.
\newblock \bibinfo{journal}{\emph{Proc. {VLDB} Endow.}} \bibinfo{volume}{17}, \bibinfo{number}{5} (\bibinfo{year}{2024}), \bibinfo{pages}{980--993}.
\newblock


\end{thebibliography}


\end{document}